\documentclass[a4paper,11pt]{article}
\pdfoutput=1 % if your are submitting a pdflatex (i.e. if you have
\usepackage{jcappub} % for details on the use of the package, please
\usepackage[T1]{fontenc} % if needed
\usepackage[mathscr]{eucal}

\usepackage{placeins}
\usepackage{subcaption}
\usepackage{colordvi} % for color text
\font\bigastfont=cmr10 scaled \magstep 3
\def\bdot{\hbox{\bigastfont .}}
\newcommand{\dotaverage}[1]{\left\langle #1 \right\rangle^{\bdot}_\cD}

\newcommand{\cD}{{\cal D}}
\newcommand{\average}[1]{\left\langle #1 \right\rangle_\cD}

\newcommand{\ini}{{\rm ini}}

\title{\boldmath Growth rate of spherical voids with non-comoving Dark Matter and Baryons}

\author[a,b,d]{Fernando A. Piza\~na,}
\author[a,1]{Juan Carlos Hidalgo,\note{Corresponding author.}}
\author[c]{Ismael Delgado Gaspar}
\author[b]{Roberto A. Sussman}

\affiliation[a]{Instituto de Ciencias F\'isicas (ICF-UNAM), Universidad Nacional Aut\'onoma de M\'exico, Avenida Universidad S/N, Champila, Cuernavaca, 62210, Morelos, M\'exico}
\affiliation[b]{Instituto de Ciencias  Nucleares (ICN-UNAM), Universidad Nacional Aut\'onoma de M\'exico, Circuito Exterior s/n, C.U., Coyoac\'an, 04510, Ciudad de M\'exico, M\'exico}
\affiliation[c]{Department of Fundamental Research, National Centre for Nuclear Research, Pasteura 7, 02--093 Warsaw, Poland}
\affiliation[d]{Institute of Astronomy, Faculty of Physics, Astronomy and Informatics,
Nicolaus Copernicus University, Grudziadzka 5, 87-100 Toru\'n, Poland}

\emailAdd{klesto92@ciencias.unam.mx}
\emailAdd{hidalgo@icf.unam.mx}
\emailAdd{Ismael.DelgadoGaspar@ncbj.gov.pl}
\emailAdd{sussman@nucleares.unam.mx}

\abstract{We present numerical solutions to Einstein's equations describing large spherical cosmic voids constituted by two components; dark matter and baryons, with a non-vanishing initial relative velocity, in an asymptotically homogeneous background compatible with the $\Lambda$CDM concordance model. We compute numerically the evolution of such configurations in the dark matter frame, with a hypothetical homogeneous distribution of baryons, but respecting the values dictated by the concordance model for the average baryon-to-dark matter density ratio. We reproduce the well known formation of overdensities at the edge of the void, and recover the Lemaitre-Tolman-Bondi solutions in the comoving limit of our simulations. We compute the average growth factor of matter fluctuations, and find that it departs significantly from the linear perturbative prescription even in the comoving case, where the non-linearity of inhomogeneities has an impact.}

\begin{document}
\maketitle
\flushbottom

\bibliographystyle{hunsrt}
 
\section{Introduction}
\label{sec:intro}   
    
Since the first evidence of the current accelerated expansion, many models have been proposed to introduce a dark energy component \cite{Huterer:2017buf, dePutter:2008wt, Joyce:2014kja}, and subsequently observables have been sought for probing such alternatives to a cosmological constant both at the background level \cite{SupernovaCosmologyProject:2008ojh, Joudaki:2016kym}, as well as in the statistics of structures \cite{Li:2016bis, Achitouv:2017dcm, Barros:2020bgg, deAraujo:2021cnd}. In particular, the growth function $f = d \log D / d \log a $ is an observable widely studied in the galaxy distribution \cite{1980lssu.book.....P, Lahav:1991wc} in the search for deviations from the standard theory \cite{Batista:2014uoa}. In theory, this is customarily calculated from the evolution of perturbations within the model and parametrised with the growth index $\gamma$, where $f = \Omega_m^{\gamma}$ \cite{Buenobelloso:2011sja}. The observational counterpart is constrained through the Redshift Space Distortions (RSD) manifest in the dipole and quadrupole of the galaxy power spectrum \cite{Kaiser:1987qv}. The different prescriptions for the growth function in several dark energy models has motivated the advancement of surveys of great precision in the determination of this observable \cite{DES:2022qdz, DESI:2016fyo, EUCLID:2011zbd, Euclid:2021xmh}. For example, several models propose a scale-dependence of this function, aside from the redshift dependence allowed by the above parametrisation. The experimental evidence allows for such dependence and deviations from the standard $\Lambda$CDM prescription are still open. 

In the precise characterization of this observable it is important to account for possible effects from velocity bias (see e.g. \cite{Sorrenti:2022zat}). Specifically, if the baryon field is not exactly comoving with the dark matter structures, then the relevance of a bias in the determination of the growth function demands special attention. 
Since the magnitude of such bias is largely unknown, its effect could well lie beyond a perturbative one, and nonlinear models of structure formation are required to account for its influence on the growth function. 

While it is standard to describe the evolution of cosmological structures using linear perturbations on a FLRW background, it is insufficient when  quantities such as the density contrast $\delta$ reach non-linear values. Alternative approaches for the description of non-linear phenomena are needed such as ``spherical collapse'' \cite{padmanabhan1993structure, Sheth:2003py} and inhomogeneous cosmological models. Inhomogeneous cosmological models have the advantage over linear perturbations and spherical collapse that they are exact solutions to the field equations that reduce to FLRW models at the appropriate limit \cite{AndrzejKrasinski:1997zz}. These models can be used in a variety of scenarios (see \cite{Bolejko:2009pvd} for a comprehensive discussion of examples) including the modeling of cosmological structures. Important examples of the usefulness of these types of solutions include the quasi-spherical Szekeres Class I models, where it is possible to model arrays of multiple structures arranged in a spheroidal manner \cite{Sussman:2015fwa,Sussman:2015wna,Sussman:2017otk}.

 Relevant to this study, a particular advantage of inhomogeneus cosmological solutions is the ability to describe several matter components, such as non-comoving fluids with relative velocities between them. This description can range from simple models containing a single structure \cite{DelgadoGaspar:2018uur} to more complex arrangements of structures \cite{Najera:2020jdm}. Relative velocities are important to model since peculiar velocities are measured in supernovae data which can lie within a range of $328$ - $620$ $km/s$ \cite{Sorrenti:2022zat}.
Characterizing the effect of (non-linear) velocity bias on the growth function is imperative to determine the possible values of this observable within the standard model of cosmology. Such results may help interpret observations within the $\Lambda$CDM model before considering  more exotic theories.

In this paper we look at a toy model realization of the effect of non-linear bias \cite{Yoo:2022fun} in cosmic voids, which are a suitable scenario to test gravity \cite{Pisani:2019cvo,Correa:2021lfi,Verza:2022pdc,Massara:2022lng}. With the technique deveolped in a previous paper \cite{DelgadoGaspar:2018uur} we study the late-time evolution of spherically symmetric voids with non-comoving components of dark matter and baryons. We take the dynamical system of the Einstein Field Equations (EFEs) and solve for the evolution of two fluid components out of the proper frame. We show how these components are evolved numerically from high redshifts to the present cosmic time. Our formalism is suited for inhomogeneities with arbitrary amplitude (particularly beyond the perturbative regime). We look at the growth factor $f(z)$ for departures of the averaged void and emerging overdensity, and compare it to the perturbative prescription. We find that even a small (perturbative) initial curvature inhomogeneity yields sizable differences with the linear prescription, mostly for non-comoving fluids, and to a lesser extent for the comoving case.
     
The paper is organized as follows. Sec.~\ref{Sec:1+3} reviews the 1+3 formalism of general relativity to describe general fluids. In Sec.~\ref{Sec:non-comoving} we derive the system of equations which describe the evolution of a dark matter plus baryons configuration with a non-trivial profile of relative velocity. In  Sec.~\ref{Sec:void}, we present the evolution of a large void with two non-comoving components, in an otherwise homogeneous $\Lambda$CDM universe. 
Finally in Sec.~\ref{Sec:discussion} we discuss the prospects of detection of the effect that a relative velocity brings to the discrepancy of the growth factor from its perturbative prescription.

\section{The 1+3 Formalism}
\label{Sec:1+3}

In General Relativity there are two main formalisms for the splitting of spacetime in space and time. The 3+1 formalism deals with the foliation of spacetime through a $3D$ spacelike hypersurface $\Sigma_{t_0}$ at $t=t_0$ and a timelike vector field (orthogonal to these surfaces) which defines the time evolution to find hypersurfaces $\Sigma_t$ at $t \neq t_0$. In this way, the spacetime is foliated by the different $\Sigma_t$ hypersurfaces. In contrast the 1+3 formalism threads spacetime by defining a timlike vector field, usually the 4-velocity $u^a$ {of a set of special observers}, and through it defining orthogonal spacelike hypersurfaces. We will give a brief discussion on this formalism, an in depth discussion can be found in Ref. ~\cite{ellis_maartens_maccallum_2012}.

The core of the 1+3 formalism is the 4-velocity $u^a$ (usually comoving) and the projector operator $h_{ab}$ defined as
\begin{equation}
    h_{ab} = g_{ab}+u_au_b.
\end{equation}
It is worth mentioning that this tensor works both as a projector tensor from the 4-dimensional spacetime to the $3D$ hypersurfaces as well as a metric on those hypersurfaces. Having these quantities one can split the covariant derivative of $u^a$ in terms of its irreducible kinematical quantities
\begin{equation}
    \nabla_bu_a = \omega_{ab}+\sigma_{ab}+\frac{1}{3}\Theta h_{ab}-\dot{u}_au_b,\label{eq:CovU}
\end{equation}
where $\omega_{ab}$ is the vorticity tensor, $\sigma_{ab}$ the shear tensor, $\Theta$ the expansion scalar, and $\dot{u}_a=u^b\nabla_bu_a$ the acceleration 1-form. These quantities are used to split the Einstein equations into their 1+3 form together with the electric and magnetic parts of the Weyl tensor $E_{ac}=C_{abcd}u^bu^d$ and $\mathcal{H}_{ab}=\eta_{acd}C^{cd}_{\ \ \ be}u^e/2$ respectiveley, where $\eta_{acd}=\eta_{acde}u^e=-\sqrt{|g|}\epsilon_{acde}u^e$ is the volume form of the spacelike hypersurfaces.

Now, for a general stress-energy tensor,
\begin{equation}
T_{ab}=\rho u_au_b+2q_{(a}u_{b)}+ph_{ab}+\Pi_{ab} 
\end{equation}
with elements
 \begin{gather}
     \rho=T_{ab}u^au^b, \ p=\frac{1}{3}T_{ab}h^{ab}, \ q_a=-h_{\ a}^cT_{cb}u^b, \nonumber\\  
     \text{and} \ \Pi_{ab}=T_{\langle ab \rangle}=[h_{(a}^{\ \ c}h_{b)}^{\ d}-\frac{1}{3}h_{ab}h^{cd}]T_{cd},
 \end{gather}
we have the covariant 1+3 evolution equations, derived from the Einstein field equations:
\begin{gather}
    \dot{\rho}+(\rho +p)\Theta+\overline{\nabla}^aq_a = -2\dot{u}^aq_a-\sigma^{ab}\Pi_{ab},\label{eq:dotrho}\\
    \dot{\Theta}+\frac{1}{3}\Theta^2+4\pi G(\rho +3p)-\overline{\nabla}^a\dot{u}_a = -\sigma_{ab}\sigma^{ab}+2\omega_a\omega^a+\dot{u}_a\dot{u}^a,\\
    \dot{q}_{\langle a \rangle}+\frac{4}{3}\Theta q_a+(\rho +p)\dot{u}_a+\overline{\nabla}_ap+\overline{\nabla}^b\Pi_{ab}=-\sigma_{ab}q^b+\eta_{abc}\omega^bq^c-\dot{u}^b\Pi_{ab},\\
    \dot{\omega}_{\langle a \rangle}+\frac{2}{3}\Theta \omega_a+\frac{1}{2}\mathrm{curl} \ \dot{u}_a=\sigma_{ab}\omega^b,\\
    \dot{\sigma}_{\langle ab \rangle}+\frac{2}{3}\Theta \sigma_{ab}+E_{ab}-4\pi G\Pi_{ab}-\overline{\nabla}_{\langle a}\dot{u}_{b \rangle}=-\sigma_{c \langle a}\sigma_{b \rangle}^{\ \ c}-\omega_{\langle a}\omega_{b \rangle}+\dot{u}_{\langle a}\dot{u}_{b \rangle},\\
\begin{aligned}
    \dot{E}_{ab} + \Theta E_{ab}- \mathrm{curl} \ \mathcal{H}_{ab}+4\pi G&\big[(\rho+p)\sigma_{ab}+\dot{\Pi}_{\langle ab \rangle}+\frac{1}{4}\Theta \Pi_{ab}+\overline{\nabla}_{\langle a}q_{b \rangle}\big] =\\
     -8\pi G \dot{u}_{\langle a}q_{b \rangle}+2\dot{u}^c\eta_{cd(a}^{\vphantom{a}}\mathcal{H}_{b)}^{\ d}+3\sigma_{c\langle a}^{\vphantom{a}}E_{b \rangle}^{\ \ c}-&\omega^c\eta_{cd(a}^{\vphantom{a}}E_{b)}^{\ \ d}-4\pi G\big(\sigma^c_{\ \langle a}\Pi_{b \rangle c}^{\vphantom{a}}-\omega^c\eta_{cd(a}^{\vphantom{a}}\Pi_{b)}^{\ \ d}\big),
\end{aligned}\\
\begin{aligned}
    \dot{H}_{\langle ab \rangle}+ \Theta \mathcal{H}_{ab}+\mathrm{curl} \ E_{ab}-4\pi &G\mathrm{curl} \ \Pi_{ab} = 3\sigma_{c\langle a}^{\vphantom{a}}\mathcal{H}_{b\rangle}^{\ c}-\omega^c\eta_{cd(a}^{\vphantom{a}}\mathcal{H}_{b)}^{\ d}-\\
    2\dot{u}^c\eta_{cd(a}^{\vphantom{a}}E_{b)}^{\ d}+4\pi &G\big(\sigma^{c}_{\ (a}\eta^{\vphantom{a}}_{b)cd}q^d-3\omega_{\langle a}q_{b \rangle}\big),
\end{aligned}
\end{gather}
the above system is complemented by the constraint equations derived from the Ricci identities
\begin{gather}
    \overline{\nabla}^a\omega_a=\dot{u}^a\omega_a,\\
    \overline{\nabla}^b\sigma_{ab}-\mathrm{curl} \ \omega_a-\frac{2}{3}\overline{\nabla}_a\Theta+8\pi Gq_a=-2\eta_{abc}\omega^b\dot{u}^c,\\
    \mathrm{curl} \ \sigma_{ab}+\overline{\nabla}_{\langle a}\omega_{b \rangle}-\mathcal{H}_{ab}=-2\dot{u}_{\langle a}\omega_{b \rangle},\\
    \overline{\nabla}^bE_{ab}+\frac{4\pi G}{3}\big(\overline{\nabla}^b\Pi_{ab}-2\overline{\nabla}_a\rho+2\Theta q_a \big) = \nonumber\\
    \eta_{abc}\sigma^b_{\ d}H^{cd}-3\mathcal{H}_{ab}\omega^b+4\pi G \big(\sigma_{ab}+3\eta_{abc}\omega^c\big)q^b,\\
    \overline{\nabla}^b\mathcal{H}_{ab}+4\pi G\big[\mathrm{curl}\ q_a-2(\rho+p)\omega_a \big] = \nonumber \\
    3E_{ab}\omega^b-\eta_{abc}\sigma^b_{\ d}E^{cd}-4\pi G \big(\eta_{abc}\sigma^b_{\ d}\Pi^{cd}+\Pi_{ab}\omega^b\big),\\
    ^{\tiny{(3)}}R_{\langle ab \rangle}=E_{ab}+4\pi G\Pi_{ab}-\frac{1}{3}\Theta(\sigma_{ab}+\omega_{ab})+\sigma_{c\langle a}^{\vphantom{a}}\sigma_{b \rangle}^{\ \ c}+\omega_{c\langle a}^{\vphantom{a}}\omega_{b \rangle}^{\ \ c}-2\sigma_{c[a}^{\vphantom{a}}\omega_{b]}^{\ \ c},\\
    ^{\tiny{(3)}}R=16\pi G\rho-\frac{2}{3}\Theta^2+\sigma_{ab}\sigma^{ab}-\omega_a\omega^a. \label{eq:ConstHam}
\end{gather}
Here $\overline{\nabla}_aS_{b}^{\ c}=h_b^{\ d}h_b^{\ e}h_f^{\ c}\nabla_dS_e^{\ f}$ is the projected covariant derivative, $\dot{S}_{ab} = u^c\nabla_cS_{ab}$ is the derivative projected along the flow of the 4-velocity and $\mathrm{curl}\left[\ S_{ab}\right] = \eta_{cd(a}^{\vphantom{a}}\overline{\nabla}^cS^{\ d}_{b)}$.

\section{Spherically Symmetric Equations for Non-Comoving fluids}
\label{Sec:non-comoving}

We will now move to the problem in question: The evolution of a spherically symmetric void constituted by two non-comoving, self-gravitating pressureless fluids. We first present in detail the equations that describe the evolution of our configuration. We first show how a relative velocity between fluid components turns the matter fields oberver-dependent and secondly, we use a spherically symmetric metric to obtain the explicit equations for the two non-comoving dust components. 
\subsection{Relative velocity between fluids}

Between two or more non-comoving fluids, there is a transformation between frames depending on the different relative velocities. For relative velocities $v^a_{\textrm{\tiny{I}}}$ between a component with 4-velocity $u^a$ and the components with 4-velocity $u_{\textrm{\tiny{I}}}^a$, we use the inverse transformation given by 
\begin{equation}
    u^a=\gamma_{\textrm{\tiny{I}}}(u_{\textrm{\tiny{I}}}^a+\hat{v}^a_{\textrm{\tiny{I}}}), \quad \mathrm{with}\quad \hat{v}^a_{\textrm{\tiny{I}}}=-\gamma_{\textrm{\tiny{I}}}(v_{\textrm{\tiny{I}}}^a+v^2_{\textrm{\tiny{I}}}u^a),
\end{equation}
where $g_{ab}u^a_{\textrm{\tiny{I}}}v^a_{\textrm{\tiny{I}}}=0$, $v^2_{\textrm{\tiny{I}}}=g_{ab}v^a_{\textrm{\tiny{I}}}v^b_{\textrm{\tiny{I}}}$, $\gamma_{\textrm{\tiny{I}}} = (1-v^2_{\textrm{\tiny{I}}})^{(-1/2)}$, and the subindex $I$ denotes the $I$-th fluid component with a 4-velocity $u^a_{\textrm{\tiny{I}}}$. This results in the following components of a transformed stress-energy tensor for multiple fluids with a barotropic equation of state
\begin{gather}
\begin{aligned}
    \rho_{\textrm{\tiny{I}}}=\gamma^2_{\textrm{\tiny{I}}}(1+\omega_{\textrm{\tiny{I}}}v^2_{\rm \tiny{I}})\rho^*_{\textrm{\tiny{I}}}, \ \ p_{\textrm{\tiny{I}}}=\big[ \omega_{\textrm{\tiny{I}}}+\frac{1}{3}\gamma^2_{\textrm{\tiny{I}}}v^2_{\textrm{\tiny{I}}}(1+\omega_{\textrm{\tiny{I}}}) \big]\rho^*_{\textrm{\tiny{I}}}, \\
    q^a_{\textrm{\tiny{I}}}=\gamma^2_{\textrm{\tiny{I}}}(1+\omega_{\textrm{\tiny{I}}})\rho^*_{\textrm{\tiny{I}}}v^a_{\textrm{\tiny{I}}}, \ \ \Pi^{ab}_{\textrm{\tiny{I}}}=\gamma^2_{\textrm{\tiny{I}}}(1+\omega_{\textrm{\tiny{I}}})\rho^*_{\textrm{\tiny{I}}}v^{\langle a}_{\textrm{\tiny{I}}}v^{b \rangle}_{\textrm{\tiny{I}}}.
\end{aligned}\label{eq:veltransf}
\end{gather}
The $*$ index in $\rho^*_{\textrm{\tiny{I}}}$ denotes that it is measured in its respective proper frame, and   $w_{\textrm{\tiny{I}}} = p^*_{\textrm{\tiny{I}}} / \rho^*_{\textrm{\tiny{I}}}$.

\subsection{Spherically Symmetric Inhomogeneous Equations}

We study a spherical void density configuration of two distinct dust fluid components, Cold Dark Matter (CDM) and baryons, in a $\Lambda$CDM background, so we consider the following inhomogeneus metric
\begin{equation}
    \mathrm{d}s^2=-N(t,r)^2\mathrm{d}t^2+B(t,r)^2\mathrm{d}r^2+Y(t,r)^2(\mathrm{d}\theta^2+\sin^2{(\theta)}\mathrm{d}\varphi^2).\label{eq:metricainhom}
\end{equation}
For this metric, the 4-velocity is given by $u^{a}=\delta^a_{\ t}/N(t,r)$. The covariant objects (Hubble scalar, shear tensor and the electric part of the Weyl tensor) are given, in terms of scalar functions and the basis tensor for spacelike, symmetric and trace free Petrov D spacetimes $e^{a}_{\ b}=h^{a}_{\ b}-3n^{a}n_{b}=\text{Diag}[0,-2,1,1]$, as \cite{vanElst:1995eg}
\begin{equation}
    \sigma^{a}_{\ b}=\Sigma e^{a}_{\ b} = -\frac{Y\partial_tB-B\partial_tY}{3BNY}e^{a}_{\ b}, \ E^{a}_{\ b}=W e^{a}_{\ b}, \ \text{and} \ H=\frac{\Theta}{3}=\frac{Y\partial_tB+2B\partial_tY}{3BNY},
\end{equation}
where $n_{a}=\sqrt{g_{rr}}\delta^r_{\ a}$, $\Sigma=\Sigma(t,r)$, and $W = W(t,r)$. Note that we are choosing non-rotating fluids so $\omega_{ab}=0=\mathcal{H}_{ab}$. Aditionally, we define the relative velocity, the energy flux and the anisotropic stress tensor, also in terms of scalar functions, as
\begin{equation}
    v^{a}=\frac{V(t,r)}{B(t,r)^2}\delta^a_{\ r}, \quad q^{a}=\frac{Q(t,r)}{B(t,r)^2}\delta^a_{\ r}, \quad \text{and} \quad \Pi^{a}_{\ b}=\Pi(t,r) e^{a}_{\ b}.
\end{equation}
The variables can be complemented by the Misner-Sharp mass \cite{Harada:2013epa}, which yields an alternative expression for the electric Weyl scalar
\begin{gather}
    M=\frac{Y}{2}(1-g^{ab}\nabla_{a}Y\nabla_{b}Y)=\frac{Y}{2}\left(\frac{(\partial_tY)^2}{N^2}-\frac{(\partial_rY)^2}{B^2}+1\right)\label{eq:masaMS}\\
    \implies \ W=-4\pi\Pi-\frac{4}{3}\pi \rho+\frac{M}{Y^3}.
\end{gather}
Finally, by defining $\chi=\partial_rY$ and $K=1-\chi^2/B^2$, the 3-dimensional Ricci scalar $^{\tiny{(3)}}R$ and the extrinsic curvature $\mathcal{K}$ take the form
\begin{equation}
    ^{\tiny{(3)}}R=2\frac{\partial_r(KY)}{Y^2\chi} \ \implies \ \mathcal{K}=\frac{^{\tiny{(3)}}R}{6}.\label{eq:Rich3D}
\end{equation}
As stated before, we are considering two dust components representing CDM and baryons. We make the choice of CDM as the comoving frame with 4-velocity $u^{a}$ in a $\Lambda$CDM background. The thermodynamical variables for the fluids are, according to the velocity transformations \eqref{eq:veltransf}, given by
\begin{gather}
\begin{aligned}
    \rho=\rho_{\textrm{\tiny{CDM}}}+\gamma^2 \rho^*_{b}, \ \ \ p=p_b=\frac{1}{3}\gamma^2\frac{V^2}{B^2}\rho^*_b=\frac{1}{3}\frac{Q^2}{\gamma^2\rho^*_bB^2}, \\
    Q=Q_b=\gamma^2\rho^*_bV, \ \ \ \Pi=\Pi_b=-p.
\end{aligned}
\end{gather}
It is important to note that since we are considering dust components, then $N = 1$ which implies that $\dot{u}^a=0$. Finally, we take the initial Hubble scalar $H_*$ and the initial characteristic void size $l_*$ to obtain a set of dimensionless variables 
\begin{gather}
\begin{aligned}
\mathscr{Y} = \frac{Y}{l_*}, \ \ \xi = \frac{r}{l_*}, \ \ \mathscr{S} = \frac{\Sigma}{H_*}, \ \ \mathscr{W} = \frac{W}{H^2_*},\\
\mathscr{H} = \frac{H}{H_*} , \ \ \mathrm{p} = \frac{\kappa p}{3H^2_*}, \ \ \mu = \frac{\kappa \rho}{3H^2_*}, \ \ \mathscr{Q} = \frac{\kappa Q}{3H^2_*}, \ \ \mathscr{P} = \frac{\kappa \Pi}{3H^2_*},\\
\mathscr{M} = \frac{M}{H^2_*l^3_*}, \ \ \chi = \partial_{\xi} \mathscr{Y}, \ \ \mathscr{K} =\frac{\mathcal{K}}{H^2_*}, \lambda = \frac{\Lambda}{3H^2_*} \ \ \text{y} \ \ \alpha = \frac{1}{H_* l_*}.
\end{aligned}
\end{gather}
We note that in order to obtain the standard redshift $z$ we need to define a scale factor $a$. For this we take the asymptotic radial value of the metric function $B(t,r)$ since
\begin{equation}
    B(t,r \to \infty) \rightarrow a(t),
\end{equation}
where $a(t)$ is the FLRW scale factor. This also means that 
\begin{equation}
H(t,r\to \infty) \to \bar{H}(t).
\end{equation}
When we mention the redshift, it is obtained through these asymptotic values.
The resulting evolution equations are then given by
\begin{gather}
    \dot{\mathscr{Y}} = \mathscr{Y}(\mathscr{H}+\mathscr{S}),\label{eq:yringadim}\\
    \dot{B} = B(\mathscr{H}-2\mathscr{S}),\\
    \dot{\chi} = \frac{3}{2\alpha}\mathscr{Q}\mathscr{Y}+(\mathscr{H}-2\mathscr{S})\chi,\\
    \dot{\mathscr{H}} = -\mathscr{H}^2-2\mathscr{S}^2-\frac{1}{2}(\mu_{\textrm{\tiny{CDM}}}+\mu_b+3\mathrm{p})+\lambda,\\
    \dot{\mathscr{S}} = \mathscr{S}^2-2\mathscr{H}\mathscr{S}+\frac{3}{2} \mathscr{P}-\mathscr{W},\\
    \dot{\mu}_{\textrm{\tiny{CDM}}} = -3(\mu_{\textrm{\tiny{CDM}}}+\mathrm{p})\mathscr{H},\\
    \dot{\mu}_{b} = -3(\mu_b+\mathrm{p})\mathscr{H}-6\mathscr{P} \mathscr{S} - 2\alpha\frac{\mathscr{Q}\chi}{YB^2}-\frac{\alpha \partial_{\xi} \mathscr{Q}}{B^2}+\frac{\alpha \mathscr{Q}\partial_{\xi}B}{B^3},\\
    \dot{\mathscr{Q}} = -3\mathscr{H}\mathscr{Q}-\alpha\partial_{\xi}\mathrm{p}+2\alpha \partial_{\xi}\mathscr{P}+\frac{6\alpha \mathscr{P} \chi}{\mathscr{Y}},
\end{gather}
where $\dot{\Psi}=\partial_t \Psi/H_*$ for any variable $\Psi$. The system is complemented by the Hamiltonian constraint, the Weyl constraint and the Misner-Sharp mass (where equation \eqref{eq:yringadim} is plugged into \eqref{eq:masaMS})
\begin{gather}
    \mathscr{H}^2=\mu_{\textrm{\tiny{CDM}}}+\mu_b + \lambda -\mathscr{K}+\mathscr{S}^2, \\
    \mathscr{W} = -\frac{1}{2}(\mu_{\textrm{\tiny{CDM}}}+\mu_b+\lambda)-\frac{3}{2}\mathscr{P}+\frac{\mathscr{M}}{\mathscr{Y}^3},\\
    \mathscr{M} = \frac{\mathscr{Y}}{2}\left[\mathscr{Y}^2(\mathscr{H}+\mathscr{S})^2+\alpha \left(1-\frac{\chi^2}{B^2}\right)\right].
\end{gather}
The previous system of equations is equivalent to that presented in (Ref.~\cite{DelgadoGaspar:2018uur}). 

\section{A two-component spherical void}
\label{Sec:void}

We study two particular cases of a spherically symmetric central void: one with an initial extrinsic curvature profile of higher amplitude than the other. For both cases we match an asymptotic background universe representing a flat $\Lambda$CDM cosmology with fiducial values for the density and Hubble parameter taken from the Planck 2018 results \cite{Aghanim:2018eyx}\footnote{Taken from Table 2, the parameters for baryons $\Omega_bh^2$ and for dark mater $\Omega_ch^2$ together vary slightly from the total matter content $\Omega_mh^2$ in that table, we use the separate quantities as they are the ones relevant for this study.}. %\sout{These parameters get converted to the initial chosen redshift ($z=100$) via de Friedman equation for a flat FLRW background and are presented in Table \ref{tab:table1}.} \JCH{Quizás esto y la tabla sobran. }
%\begin{table}
%\caption{\label{tab:table1}Values for the background $\Lambda$CDM cosmology parameters at $z=0$ (denoted by a subindex $0$) and $z=100$ (denoted by a subindex i).}
%\begin{ruledtabular}
%\begin{tabular}{cccc}
% $H_0$ $\left[\frac{km}{sMpc}\right]$& $\Omega_{\textrm{\tiny{CDM}}0}$  & $\Omega_{b0}$ & $\Omega_{\Lambda0}= 1- \Omega_{\textrm{\tiny{CDM}}0} - \Omega_{b0}$ \\
%\hline
% $67.36$ & $0.2645$ & $0.0493$  & $0.6862$ \\ 
%\hline
 %$H_i$ $\left[\frac{km}{sMpc}\right]$ & $\Omega_{\textrm{\tiny{CDM}}i}$  & $\Omega_{bi}$ & $\Omega_{\Lambda i}= 1- \Omega_{\textrm{\tiny{CDM}}i} - \Omega_{bi}$ \\ 
%\hline
% $38301.08$ & $0.8429$ & $0.1571$ & $2.1224\times 10^{-6}\simeq 0$\\ 
%\end{tabular}
%\end{ruledtabular}
%\end{table}

The initial profiles for the velocity, curvature and density (for the CDM) are given by the following Gaussian profiles
\begin{gather}
    \begin{aligned}
        \mu_{\textrm{\tiny{CDM}}i} = \Omega_{\textrm{\tiny{CDM}}i}\left(1- 0.01 e^{-\left(\frac{r}{0.03}\right)^2} \right), \\ 
        V_i = V_c r^2 e^{-(\frac{r-0.01}{0.025})^2} \ \ \text{and} \ \ K_i=k_cr^2e^{-\left(\frac{r}{0.03}\right)^2}.
    \end{aligned}\label{eq:perfs}
\end{gather}
From the equations \eqref{eq:perfs} we take the two specific cases of differing curvature by using the values for the amplitude constant $k_c=-0.05$ and $k_c = -0.009$. The different velocities are modulated by the amplitude constant $V_c$, and we present 6 different velocities for each of the curvature profiles: $V_c = 0,0.2,0.6,1.0,1.4,1.8$ corresponding to maximum relative velocities, expressed in $~\mathrm{km/s}$, of $V_{\textrm{\tiny{max}}}=0,~28.472,~85.416,~142.361,~199.305,~256.249$ respectively --All of these representing values compatible with the peculiar velocities according to statistics from recent surveys \cite{Howlett2022:2201.03112v1}. Note that the comoving case of $V_c=0$ represents the Lemaitre-Tolman-Bondi (LTB) solution. We take the baryonic density profile as non-trivial only in the comoving case. That is: 
\begin{figure}[b!]
\begin{subfigure}{.5\textwidth}
  \centering
  \includegraphics[width=1\linewidth]{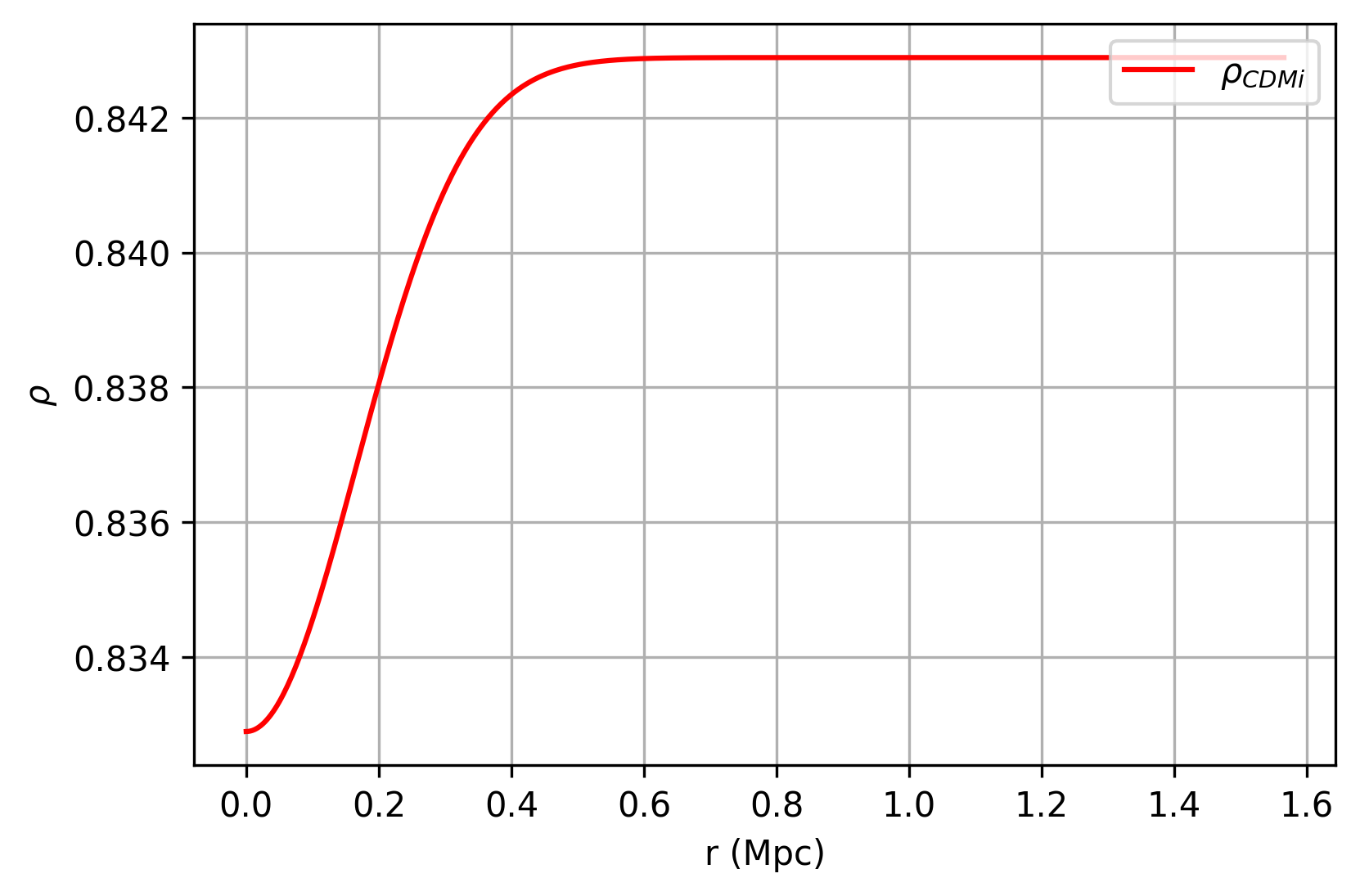}
  \label{Fig:CDMinperf}
\end{subfigure}%
\begin{subfigure}{.5\textwidth}
  \centering
  \includegraphics[width=1\linewidth]{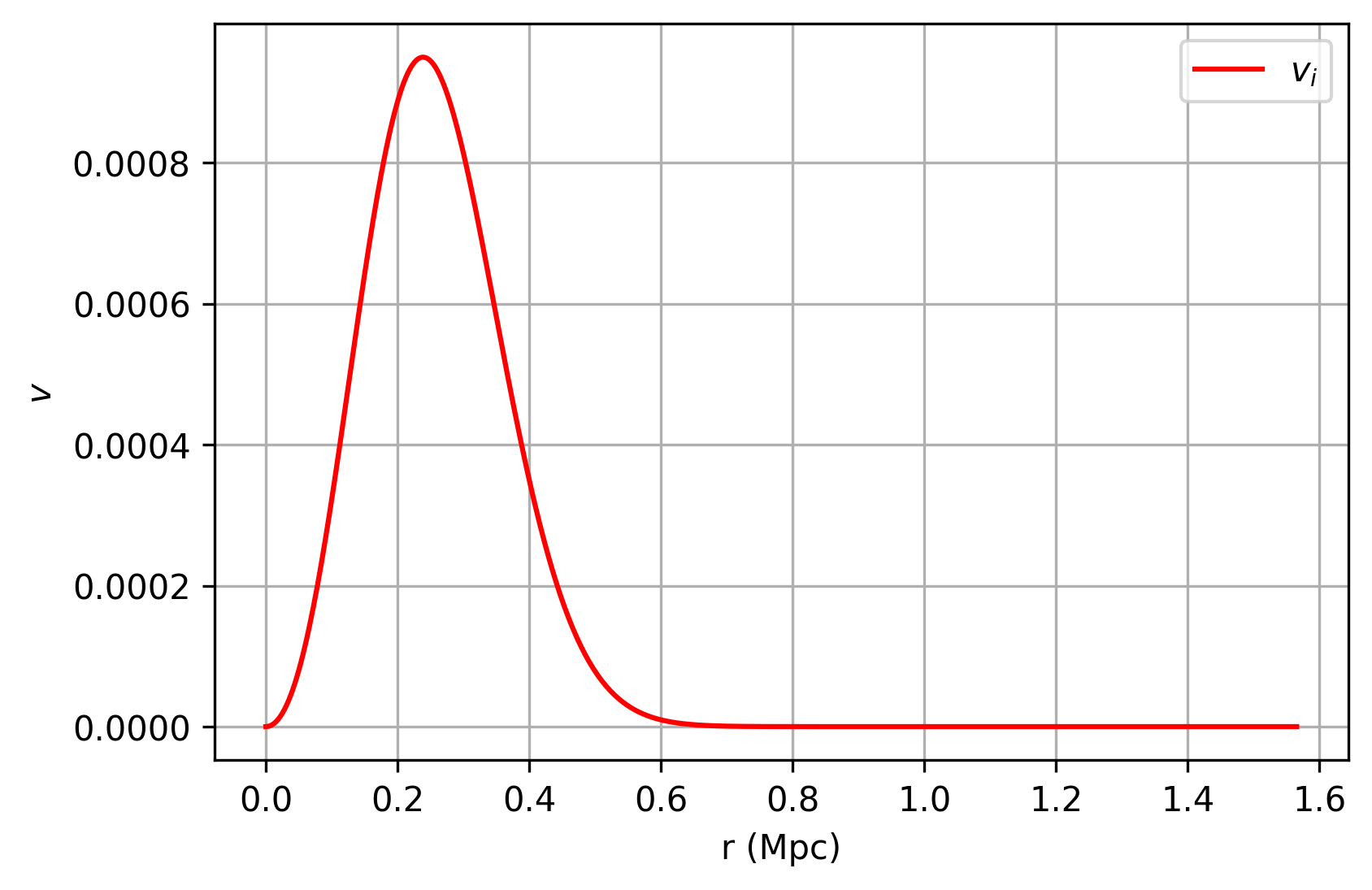}
  \label{Fig:velperf}
\end{subfigure}
\begin{subfigure}{.5\textwidth}
  \centering
  \includegraphics[width=1\linewidth]{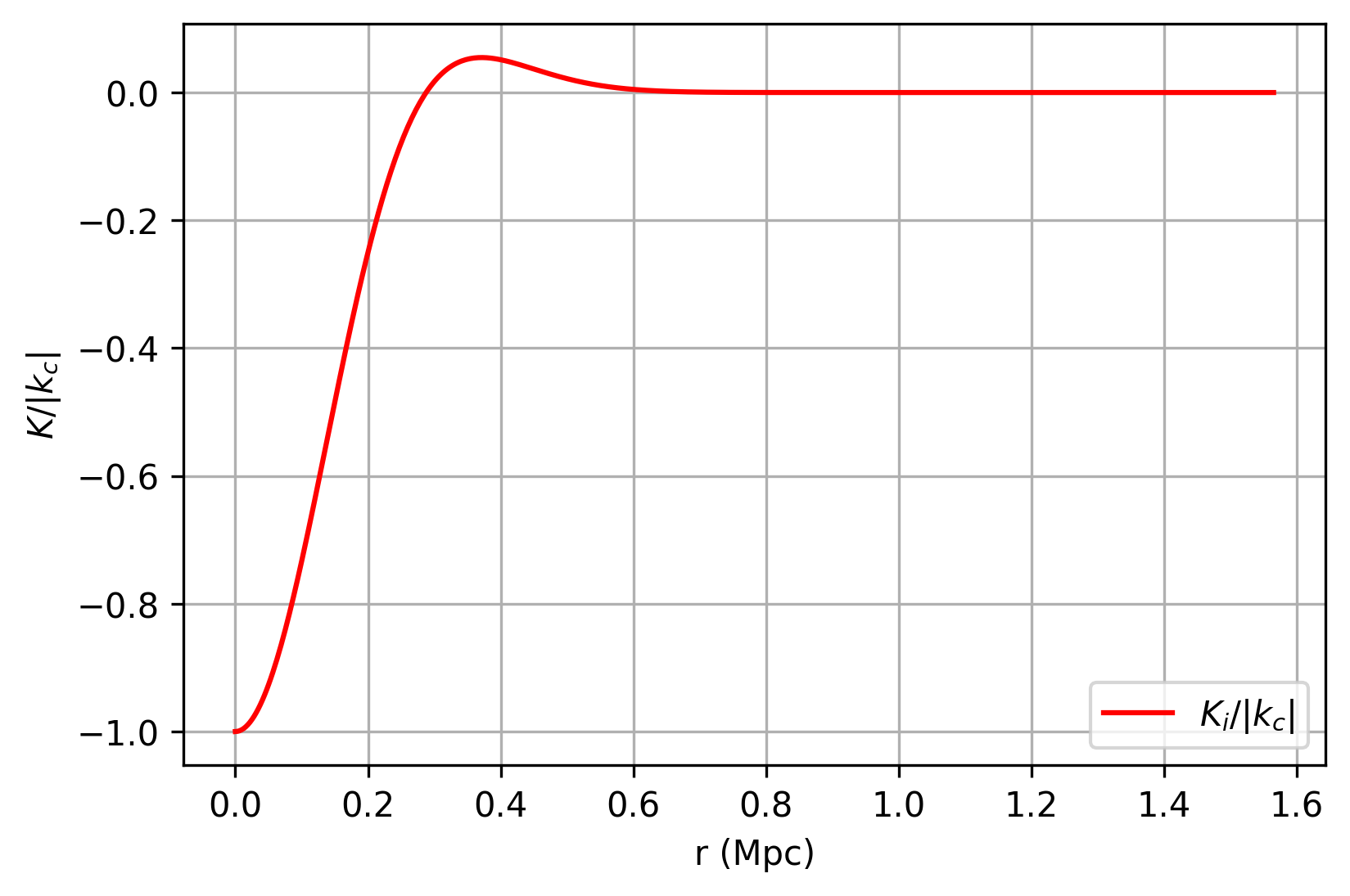}
  \label{Fig:Kinperf}
\end{subfigure}%
\caption{Initial profiles of the void configuration: the upper left is the initial Gaussian dark matter density profile, the upper right is the initial relative velocity profile for $V_c=1.8$, and the lower panel is the graph of the initial intrinsic curvature normalized by the amplitude $ \lvert k_c \rvert $.}
\label{Fig:perfsiniciales}
\end{figure}
\begin{equation}
   \mu_{bi} = 
   \begin{cases}
    \Omega_{bi}\left(1- 0.01 e^{-\left(\frac{r}{0.03}\right)^2} \right) \ \ \text{if} \ \ V_c = 0\\
    \Omega_{bi} \ \ \text{if} \ \ V_c \neq 0
    \end{cases}\label{eq:bariones}.
\end{equation}
Since the LTB case imposes a null relative velocity between baryons and dark matter, the equations naturally yield  similar radial profiles. On the other hand, that restriction is absent for a non-zero relative velocity, in which case we chose a homogeneous baryonic density profile as an initial condition (as measured in their proper frame of reference). The initial profiles for CDM, velocity and curvature are shown in Fig.~\ref{Fig:perfsiniciales}.

\begin{figure}[ht]
\begin{subfigure}{.5\textwidth}
  \centering
  \includegraphics[width=1\linewidth]{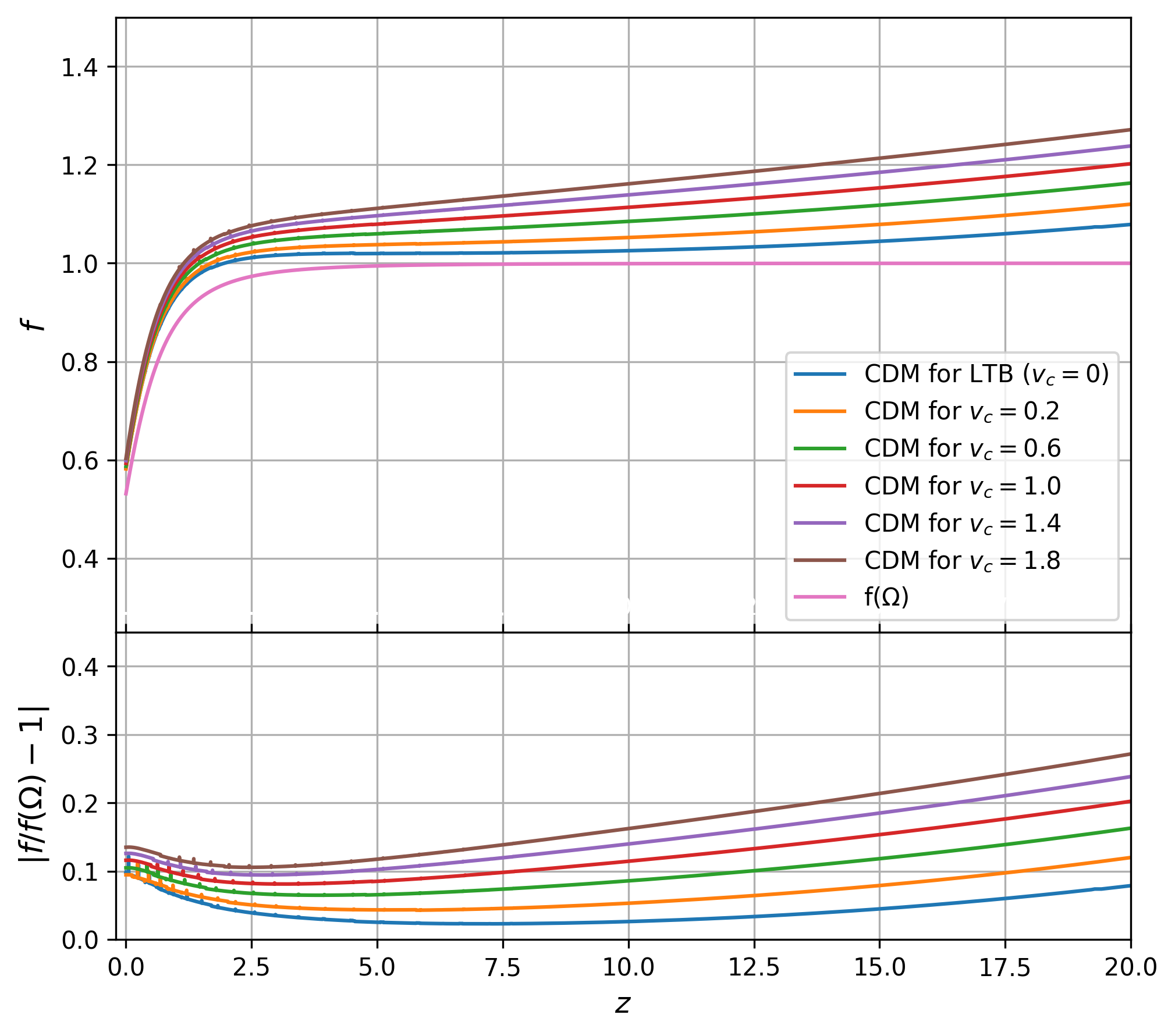}
  \label{Fig:fOmCDMOD}
\end{subfigure}%
\begin{subfigure}{.5\textwidth}
  \centering
  \includegraphics[width=1\linewidth]{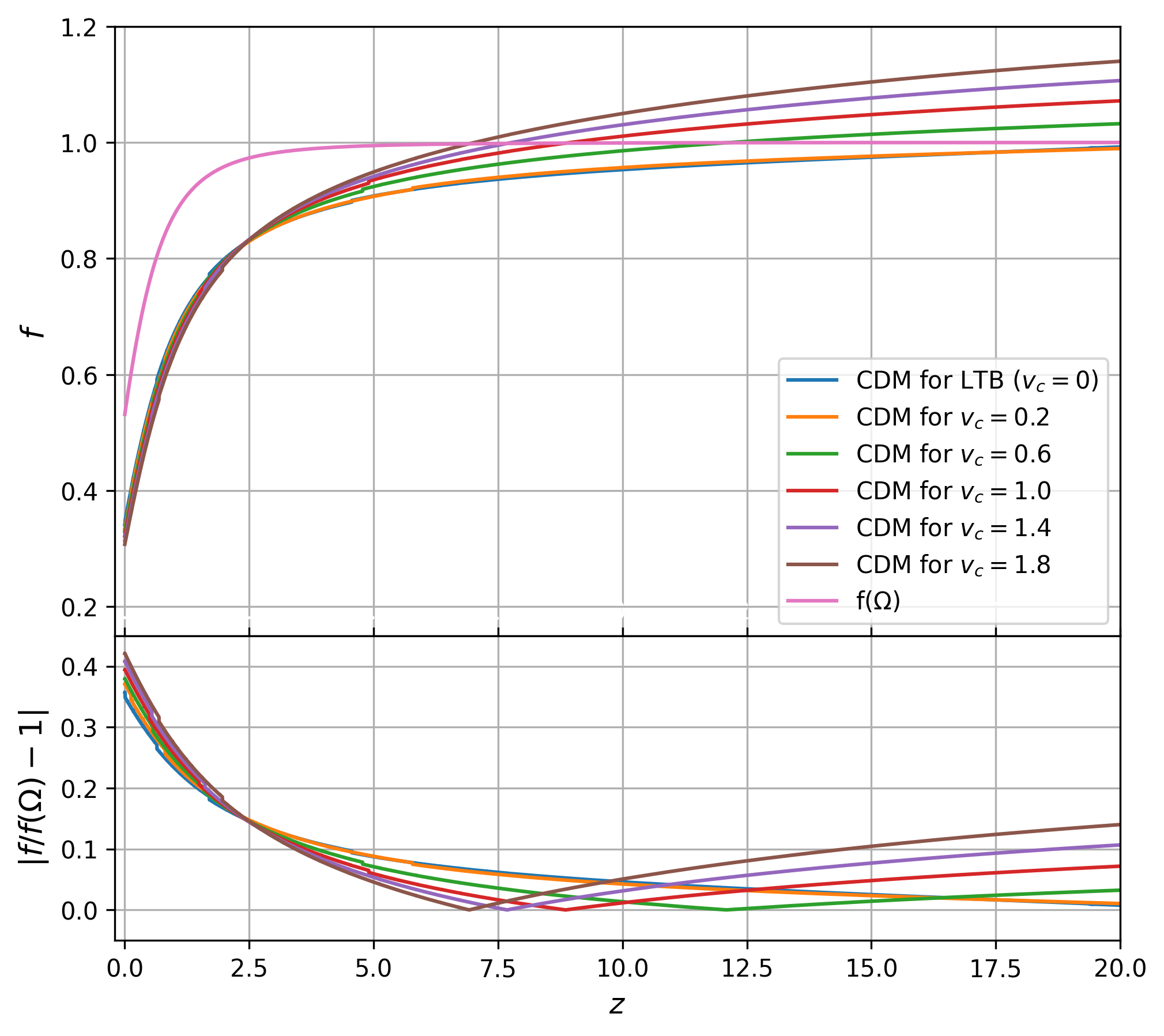}
  \label{Fig:fOmCDMV}
\end{subfigure}
\begin{subfigure}{.5\textwidth}
  \centering
  \includegraphics[width=1\linewidth]{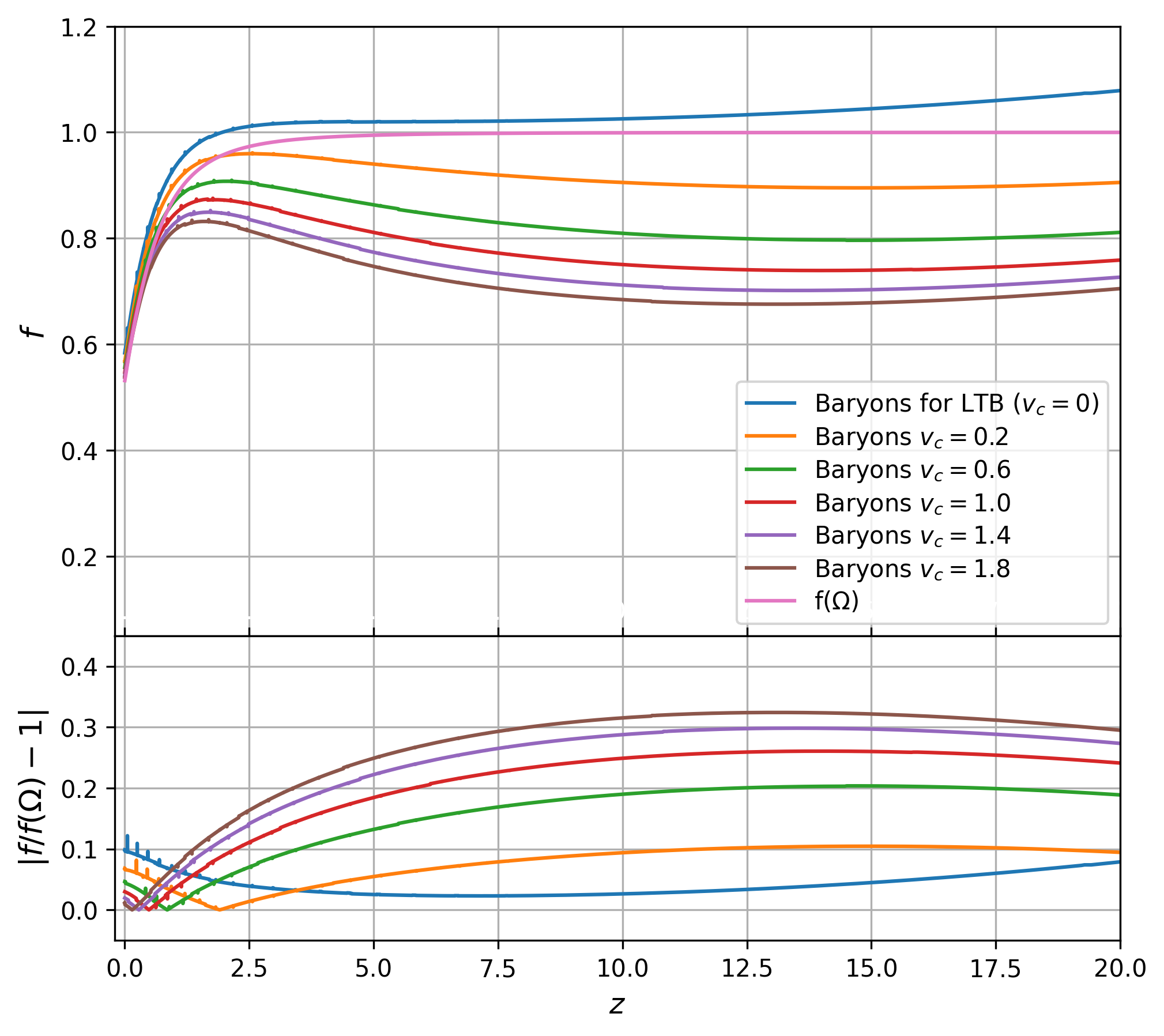}
  \label{Fig:fOmbOD}
\end{subfigure}%
\begin{subfigure}{.5\textwidth}
  \centering
  \includegraphics[width=1\linewidth]{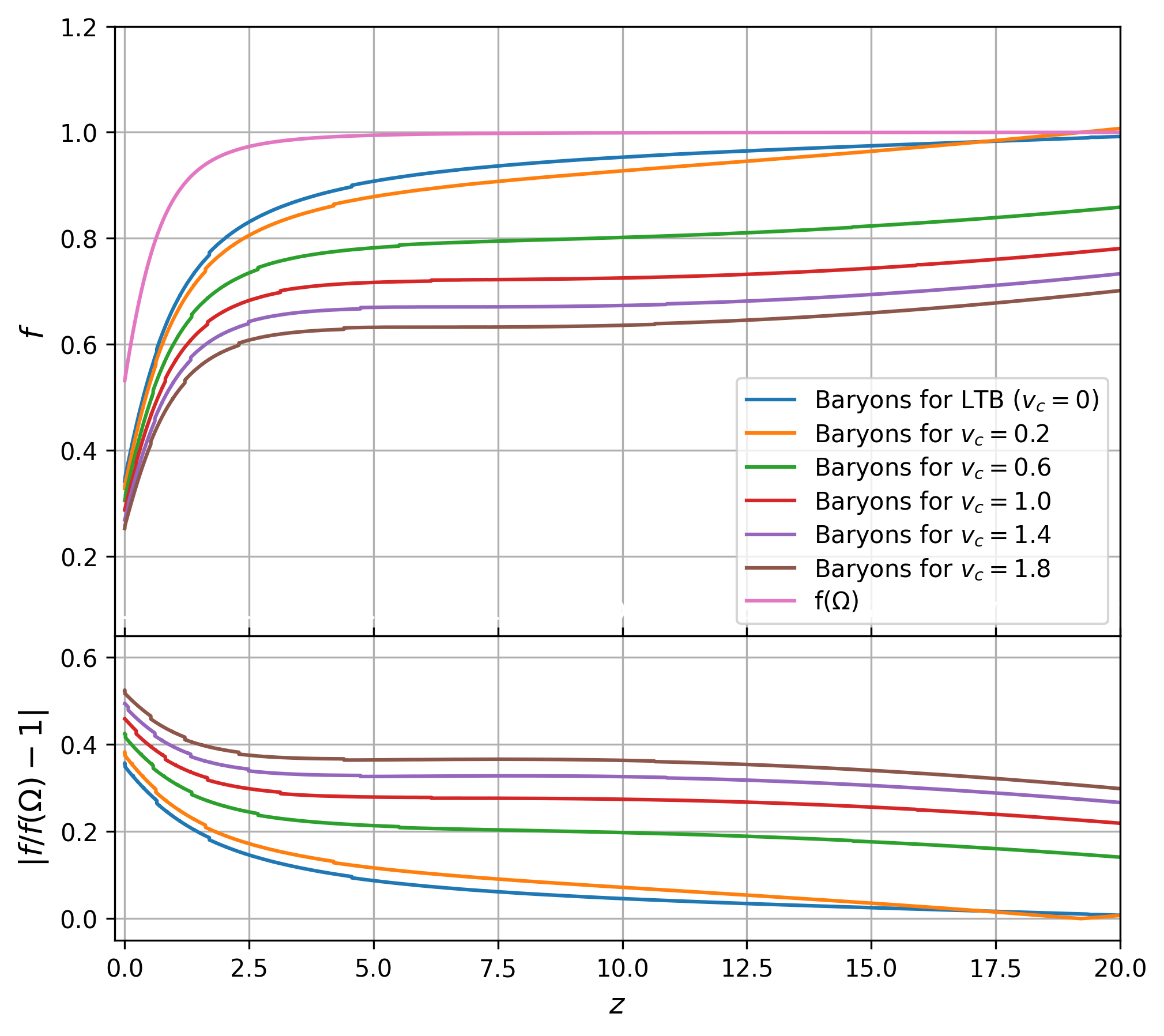}
  \label{Fig:fOmbV}
\end{subfigure}
\begin{subfigure}{.5\textwidth}
  \centering
  \includegraphics[width=1\linewidth]{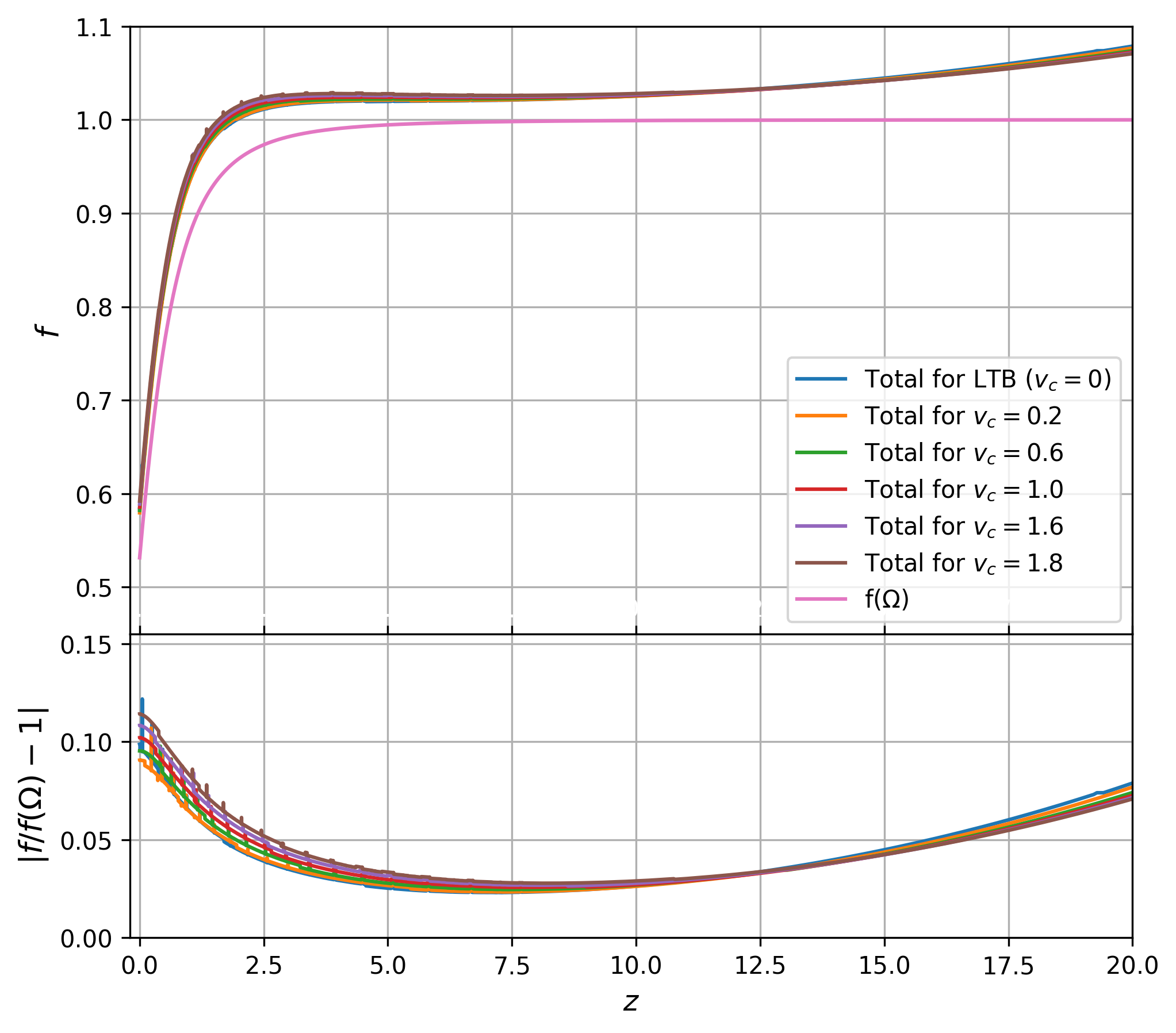}
  \label{Fig:fOmTOD}
\end{subfigure}%
\begin{subfigure}{.5\textwidth}
  \centering
  \includegraphics[width=1\linewidth]{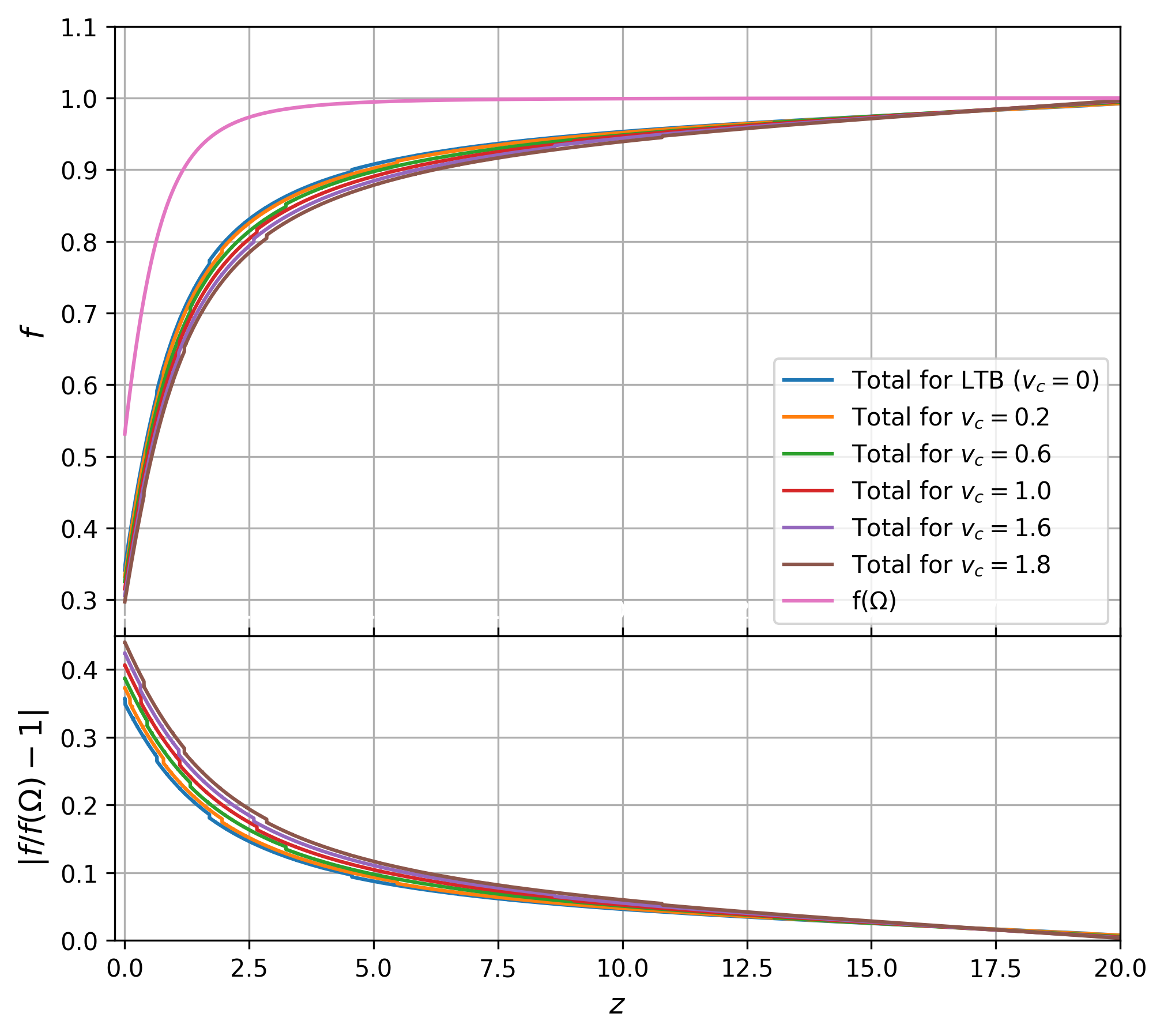}
  \label{Fig:fOmTV}
\end{subfigure}
\caption{For the higher amplitude in curvature, $k_c=-0.05$. the panels are aranged as follows: top left figure represents the growth function $f$ of CDM for multiple velocities of the over-dense ridge, top right belongs to the void region for CDM, middle left and middle right are the growth functions for the baryonic component for the over-dense ridge and the void region respectively and, finally, the bottom right and bottom left are the growth function for the total matter density contrast for the over-dense and void regions respectively. All individual figures include the percentage difference between the different growth functions for a given velocity profile and the analytic linear function.}
\label{Fig:fOm}
\end{figure}

\begin{figure}[ht]
\begin{subfigure}{.5\textwidth}
  \centering
  \includegraphics[width=1\linewidth]{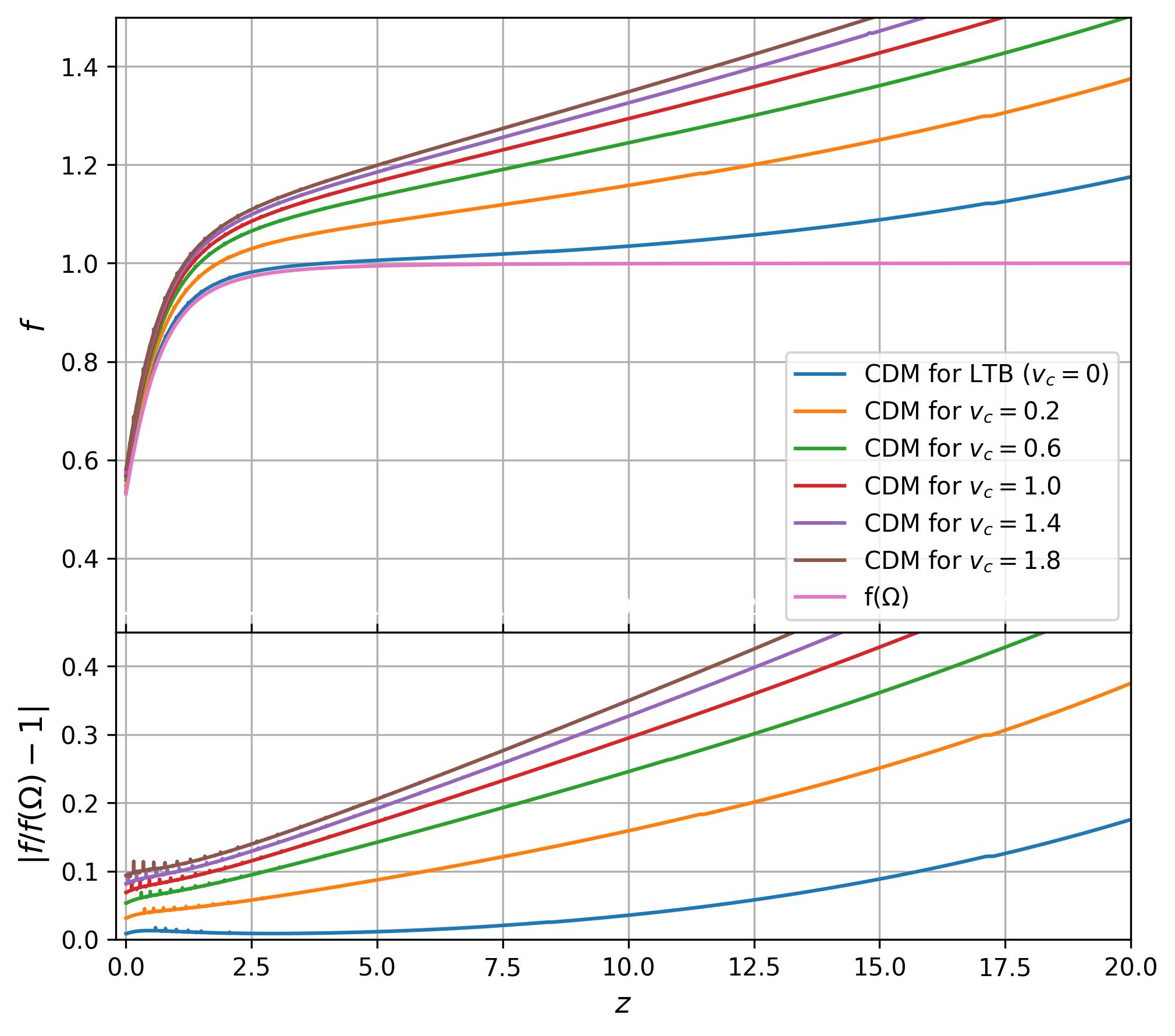}
  \label{Fig:fOmCDMODkc}
\end{subfigure}%
\begin{subfigure}{.5\textwidth}
  \centering
  \includegraphics[width=1\linewidth]{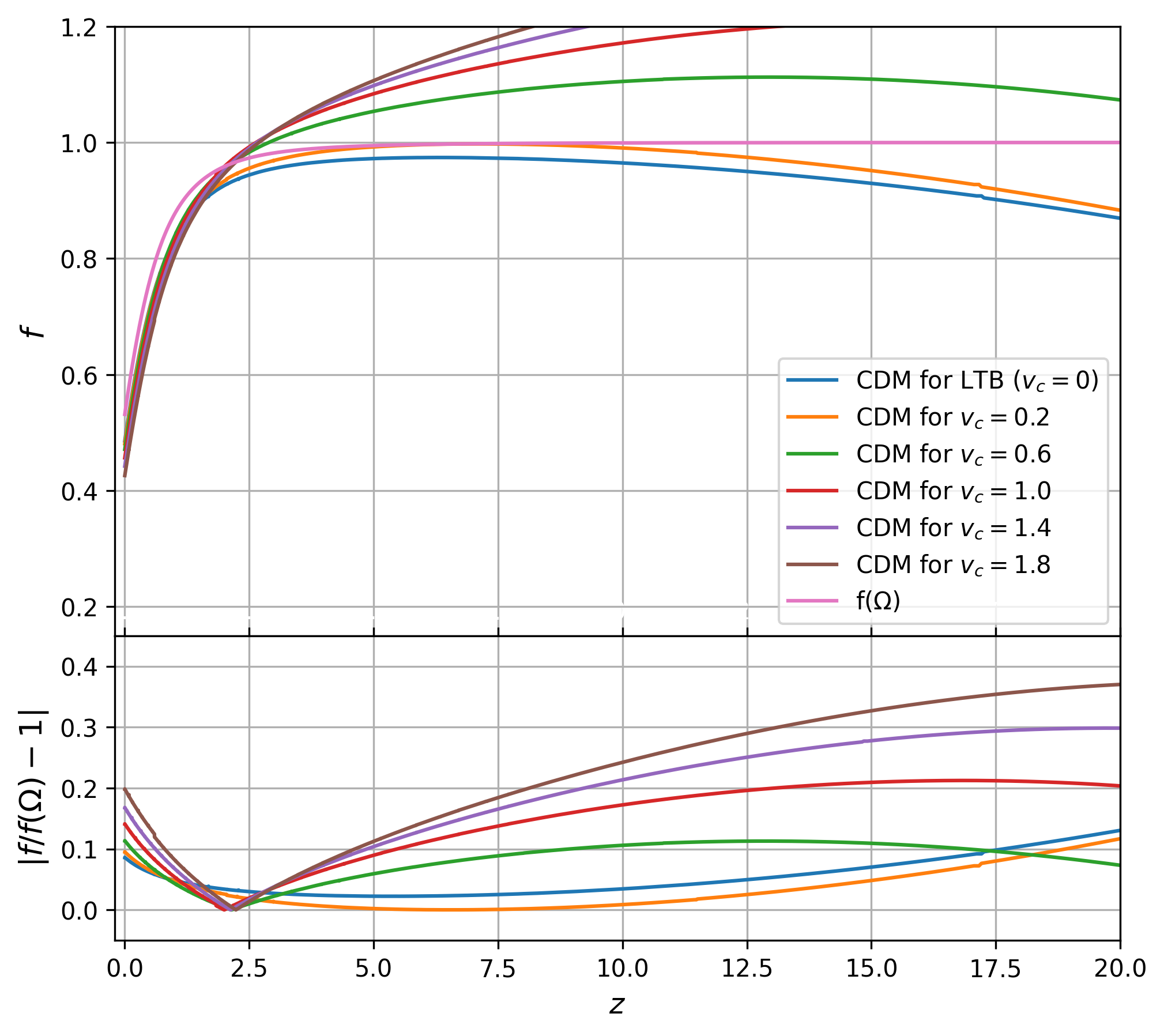}
  \label{Fig:fOmCDMVkc}
\end{subfigure}
\begin{subfigure}{.5\textwidth}
  \centering
  \includegraphics[width=1\linewidth]{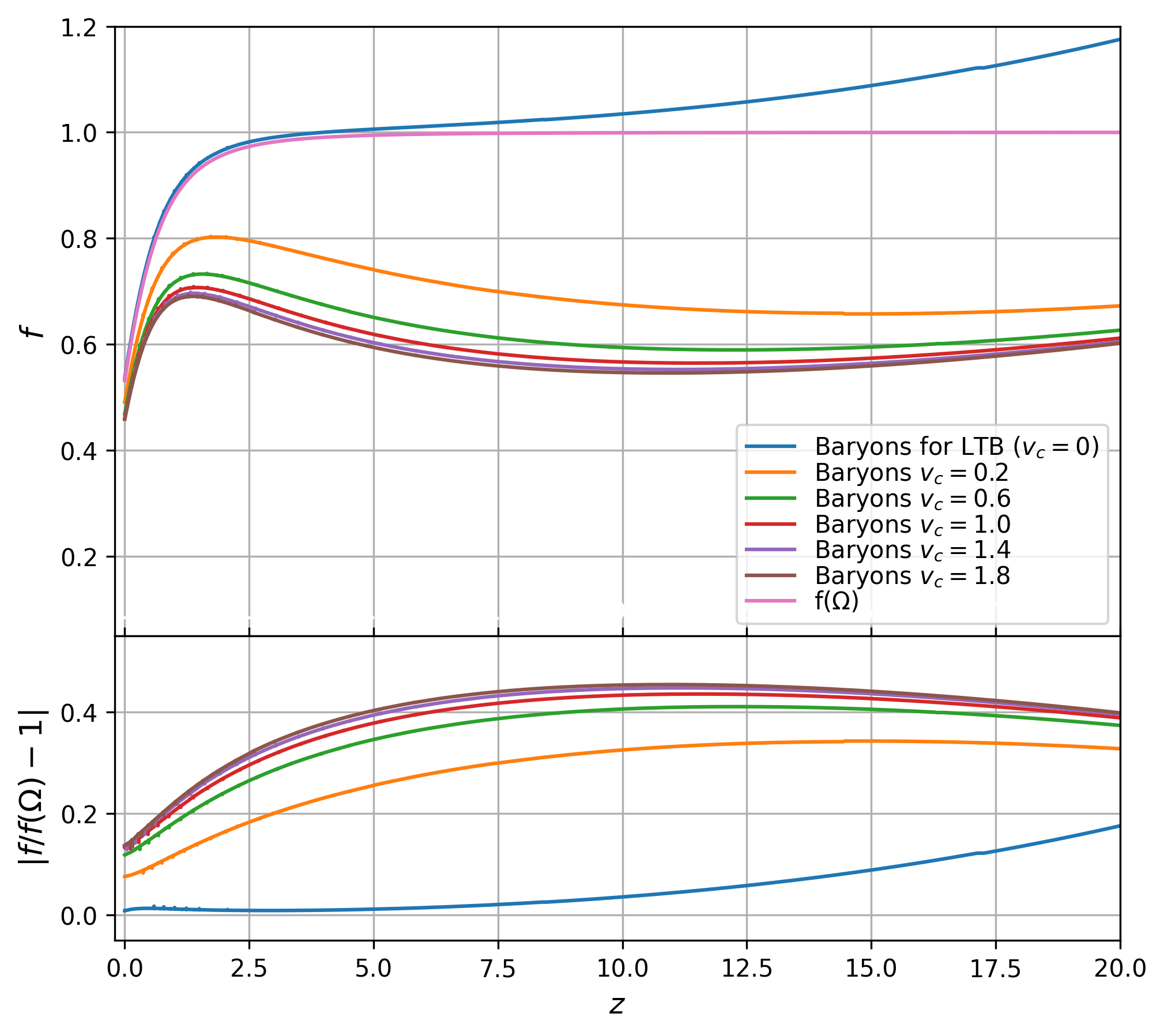}
  \label{Fig:fOmbODkc}
\end{subfigure}%
\begin{subfigure}{.5\textwidth}
  \centering
  \includegraphics[width=1\linewidth]{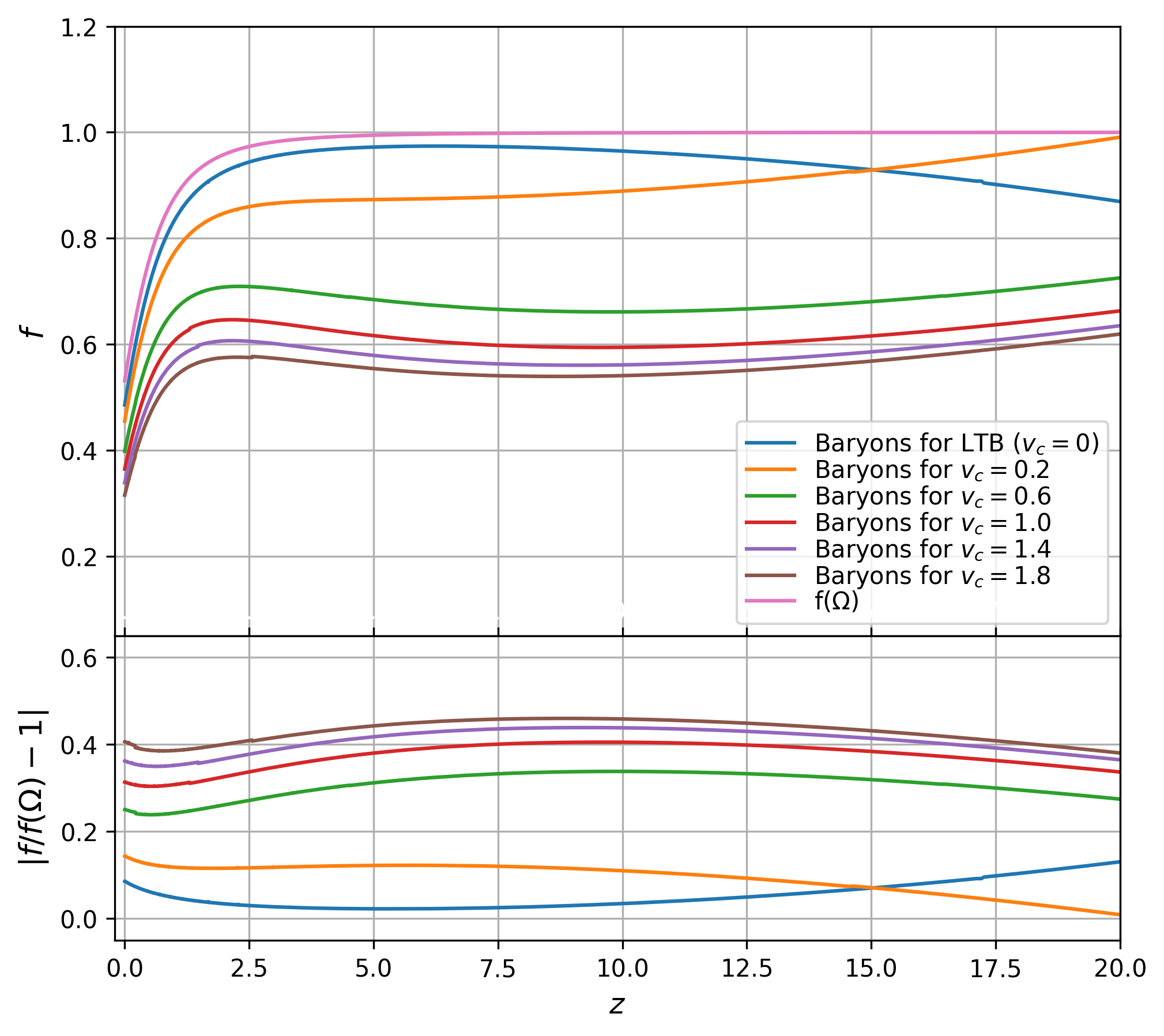}
  \label{Fig:fOmbVkc}
\end{subfigure}
\begin{subfigure}{.5\textwidth}
  \centering
  \includegraphics[width=1\linewidth]{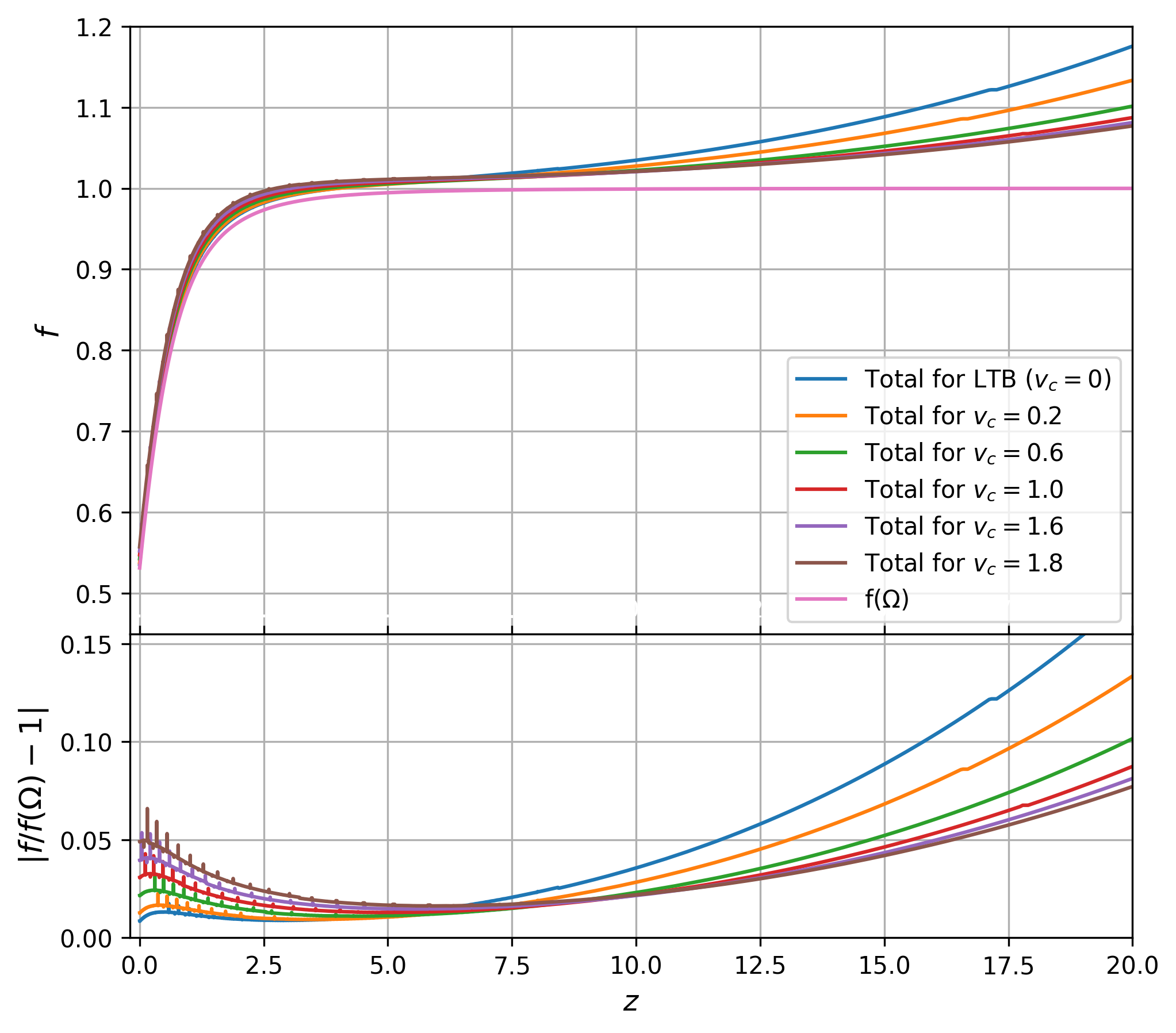}
  \label{Fig:fOmTODkc}
\end{subfigure}%
\begin{subfigure}{.5\textwidth}
  \centering
  \includegraphics[width=1\linewidth]{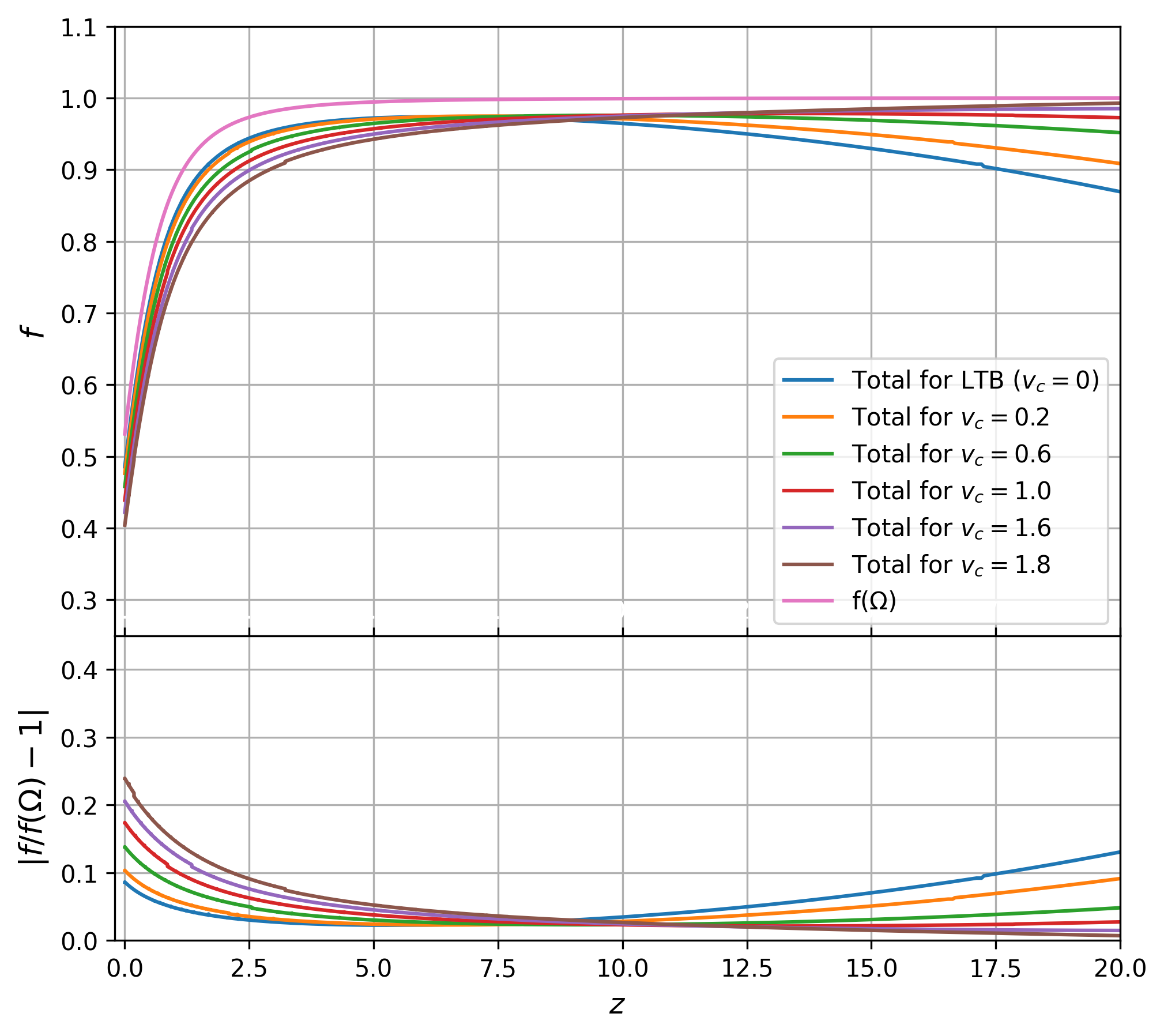}
  \label{Fig:fOmTVkc}
\end{subfigure}
\caption{The arrangement of these graphs is exactly the same as in Fig. \ref{Fig:fOm} but for the evolution of the inhomogeneity with a curvature amplitude $k_c=0.009$.}
\label{Fig:fOmkc}
\end{figure}

To analize the growth of structure, we take the definition of the growth function $f$ as given in (Ref.~\cite{peebles:80}). In the linear regime, this is defined in terms of the growing mode $D_+$ of the linear density contrast %
\begin{equation}
    f=\frac{\dot{D}_{+}}{\bar{H}D_{+}} = \frac{\mathrm{d} \log D_{+} }{\mathrm{d} \log a},
\end{equation}
In a $\Lambda$CDM background, the analytical growth function in the linear regime obeys the powerlaw (see e.g. \cite{Bruni:2013qta})
\begin{equation}
    f(\Omega_m) = \Omega_m^{6/11}.\label{eq:fomegahochgam}
\end{equation}
This prescribes the evolution, at different values of $z$, for the density contrast in the case of comoving baryons and dark matter as shown in Figs. \ref{Fig:fOm}-\ref{Fig:fOmkc}.
\begin{figure}[ht]
\begin{subfigure}{.5\textwidth}
  \centering
  \includegraphics[width=1\linewidth]{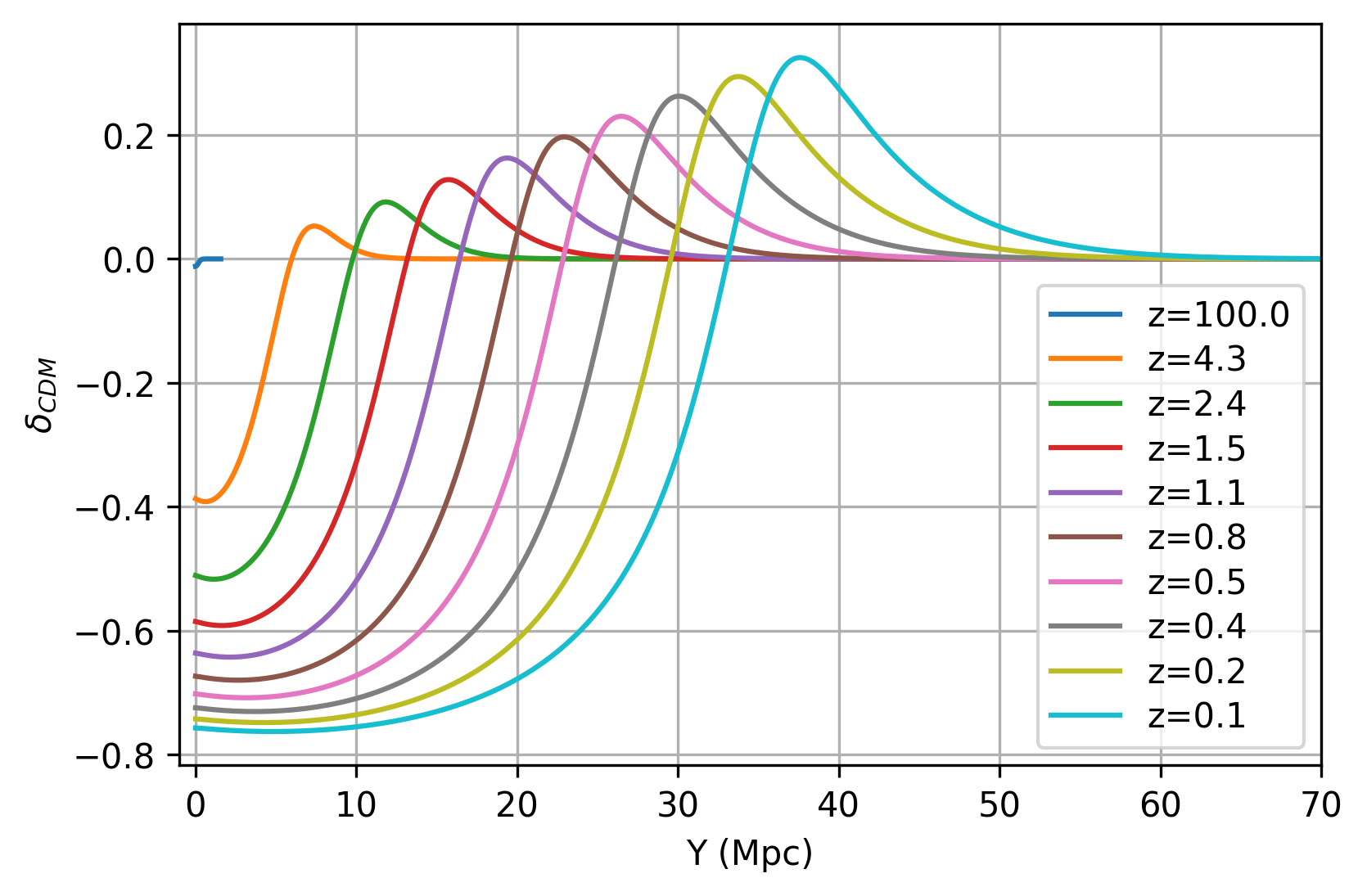}
  \label{Fig:CDMevperf}
\end{subfigure}%
\begin{subfigure}{.5\textwidth}
  \centering
  \includegraphics[width=1\linewidth]{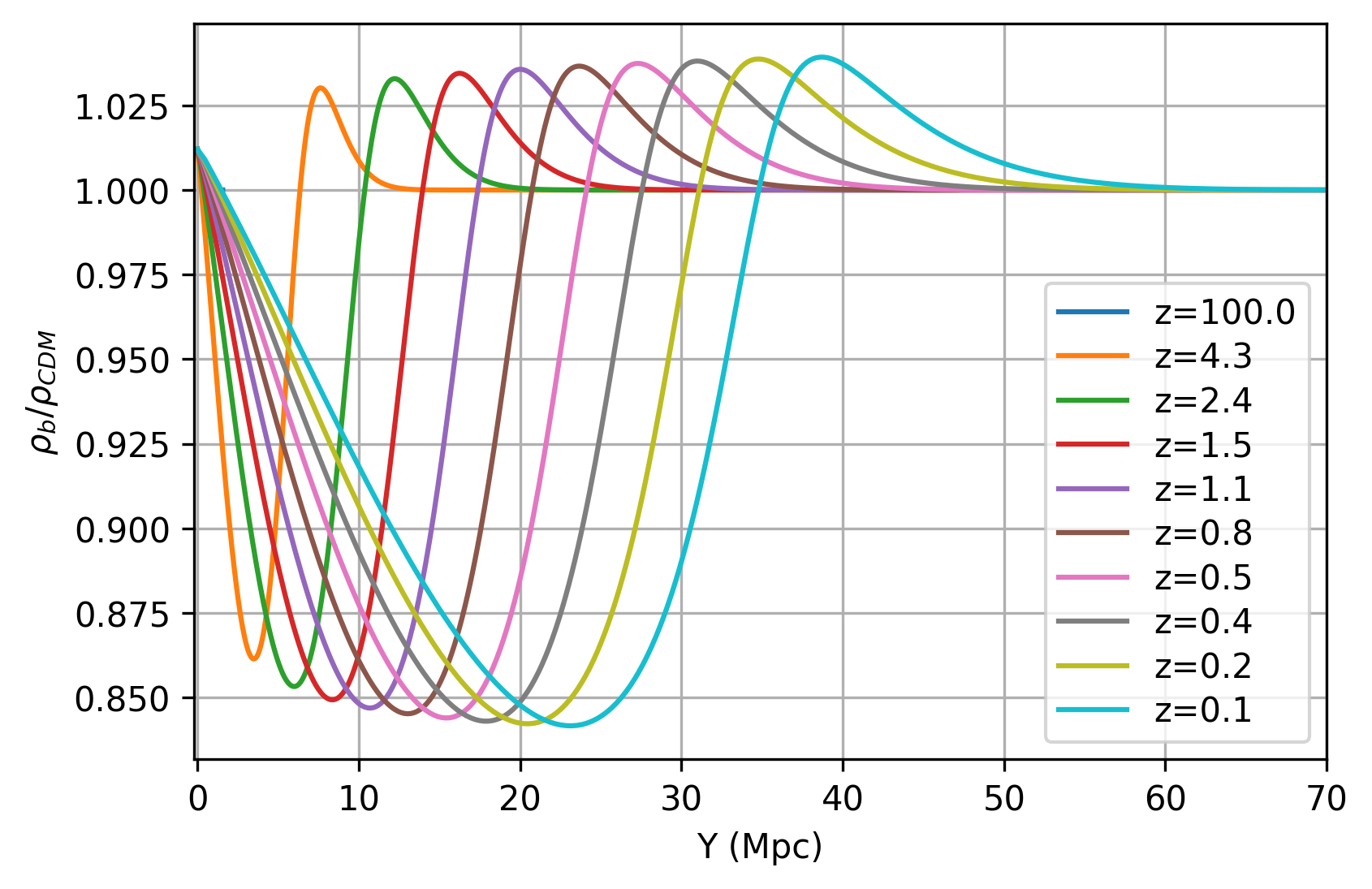}
  \label{Fig:bevperf}
\end{subfigure}
\caption{The left panel shows the evolution of Dark Matter's density contrast for different values of $z$. The right panel shows the evolution for the ratio between the two component's densities, $\rho_b/\rho_{\rm CDM}$ in the dark matter frame. Both panels correspond to the higher amplitude in curvature, $k_c=-0.05$. Both figures are the result of a relative velocity profile with $V_c=1.8$. }
\label{Fig:cdmbevperf}
\end{figure}
\begin{figure}[ht]
\begin{subfigure}{.5\textwidth}
  \centering
  \includegraphics[width=1\linewidth]{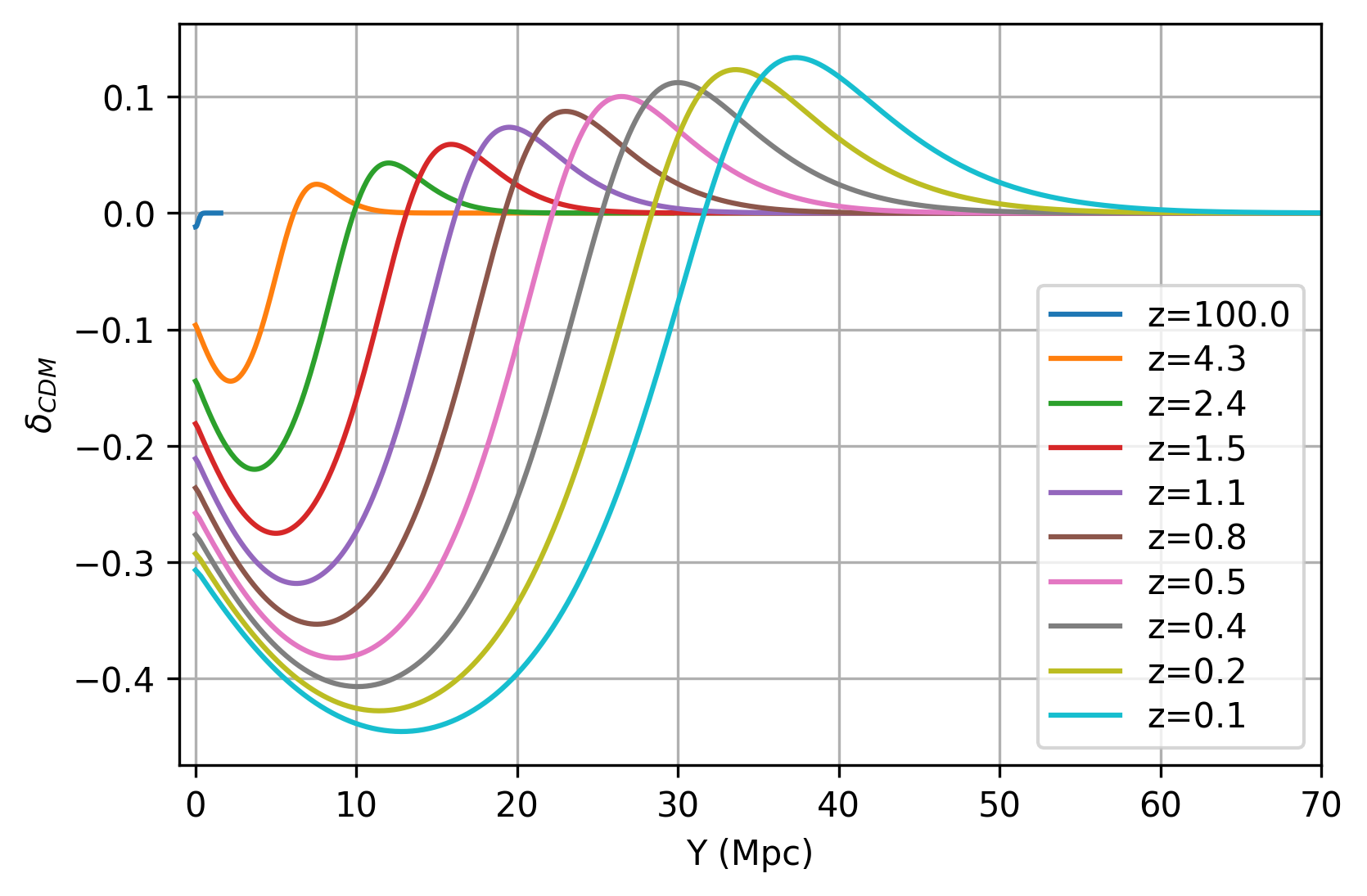}
  \label{Fig:CDMevperfkc}
\end{subfigure}%
\begin{subfigure}{.5\textwidth}
  \centering
  \includegraphics[width=1\linewidth]{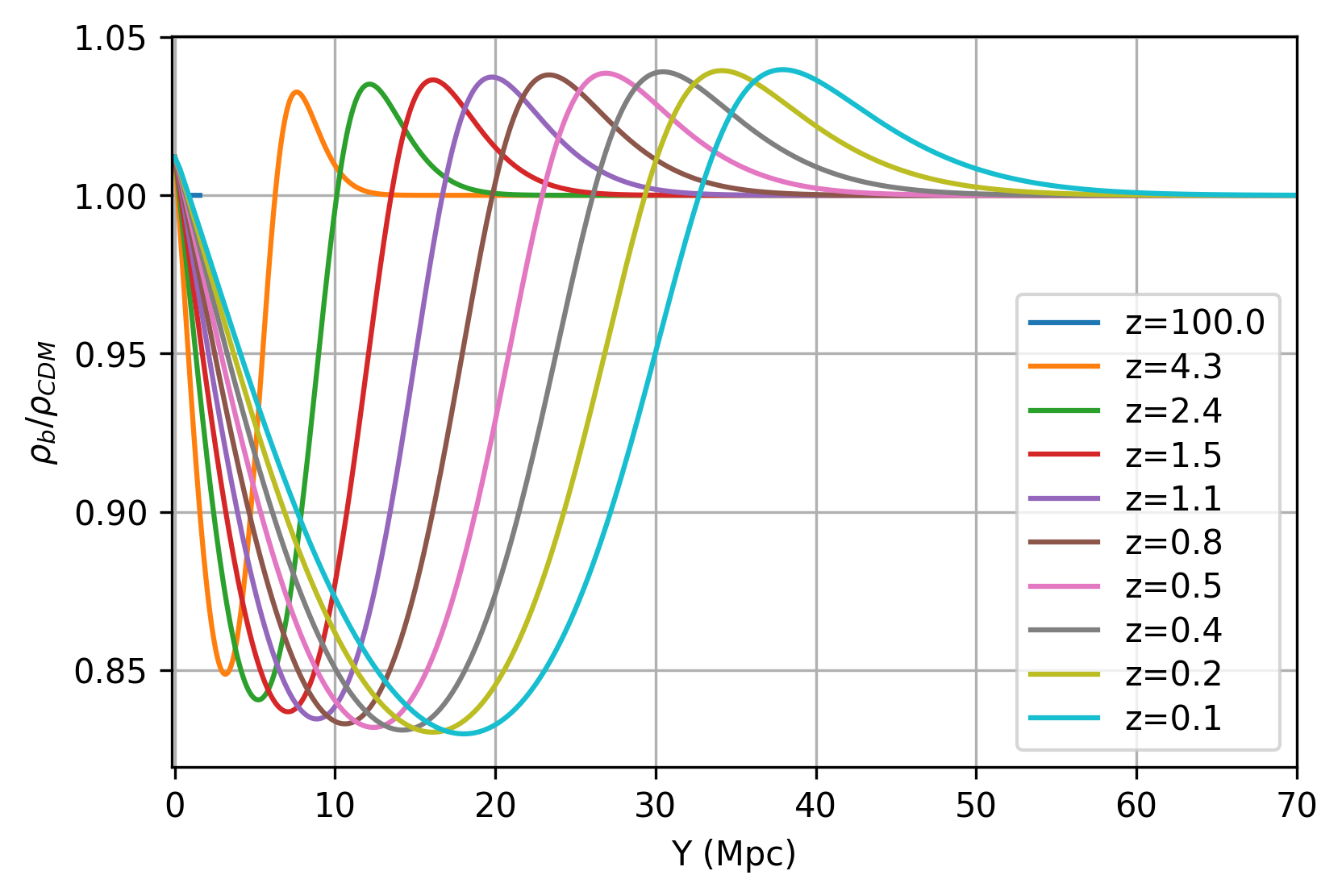}
  \label{Fig:bevperfkc}
\end{subfigure}
\caption{Like both figures in Fig. \ref{Fig:cdmbevperf} the left panel shows the evolution of Dark Matter's density contrast while the right panel corresponds to $\rho_b/\rho_{\rm CDM}$ for different values of $z$. Both figures are the result of a relative veocity profile with $V_c=1.8$ but with a curvature with an amplitude constant of $k_c=0.009$. Note the considerable amplitude difference in the profile is due to the curvature magnitude.}
\label{Fig:cdmbevperfkc}
\end{figure}
We generalize the definition of the growth function, by first defining the mean density contrast in non-linear structures. We take the average given by 
%\ismael{en la ecuacion en la derecha sugiero indicar 'volumen'. algo asi: $\frac{ \dotaverage{\delta}}{\average{\delta}}$. Luego hacer el 'dot' mas grande para que se entienda que estamos tomando en cuenta todo, incluso las variaciones del volume (por esto de la no conservation).}
%
\begin{equation}
   \average{\delta}(t) = \frac{\int_{r_1}^{r_2}\delta(t,r)Y^2B dr}{\int_{r_1}^{r_2}Y^2B dr} \implies f=\frac{\dotaverage{\delta}(t)}{\bar{H}\average{\delta}} = \frac{\mathrm{d} \log \average{\delta} }{\mathrm{d} \log B(t,r \to \infty)}.
\end{equation}
%
%\JCH{Discutir la no-conmutatividad de la derivada temporal. Evaluar la diferencia entre ambas definiciones.}`
%\JCH{Discutir acá el dominio de promediación? Es necesario advertir que la definición no es sobre coordenadas comóviles porque el fluido no es comovil y posiblemente la masa no se conserva dentro de estos radios.}
The notation $\langle \rangle_\mathcal{D}$ denotes the quantities in brackets averaged over the domain $\mathcal{D}$. Also $\dotaverage{} $ denotes the time derivative of the averaged quantity.
The motivation for adopting these definitions is presented in Appendix \ref{sec:AppA}.

%\ismael{Algunos elementos acerca de esto (es un parrafo del apendice que habia puesto):
%\\
%The physical motivation behind \eqref{GrowthFuncInh} becomes more transparent if we consider the following elements. (i) this definition reduces to \eqref{LinGrowthFunc} under the assumptions of linear perturbations ($\average{\delta^{L}} =D_+$ and $a_{\cD_H}\rightarrow a$, the Friedmann scale factor). (ii) The average extends all over the entire structure, resembling the growth of a uniform spherically symmetric overdensity detached from the expansion, as in the so-called ``spherical collapse model'' (for example, see Section 8.2 in \cite{weinberg2008cosmology}). (iii) As thus defined,  $f$ considers the variations in the domain ${\cD}$. Note that a similar definition $ f=\frac{1}{\mathcal{H}_{\cD_H}}  \average{\dot{\delta}}/\average{\delta}$ satisfies $(i)$ and $(ii)$ but fails to include the effects of changes in ${\cD}(t)$.}
%\ismael{si creen que valga la pena, podemos decir que esta definicion puede extenderse a modelos totalmente inhomogeneos no-esfericos como se indica en el apendice. Pero no es necesario :)}

\begin{figure}[ht]
\begin{subfigure}{.5\textwidth}
  \centering
  \includegraphics[width=1\linewidth]{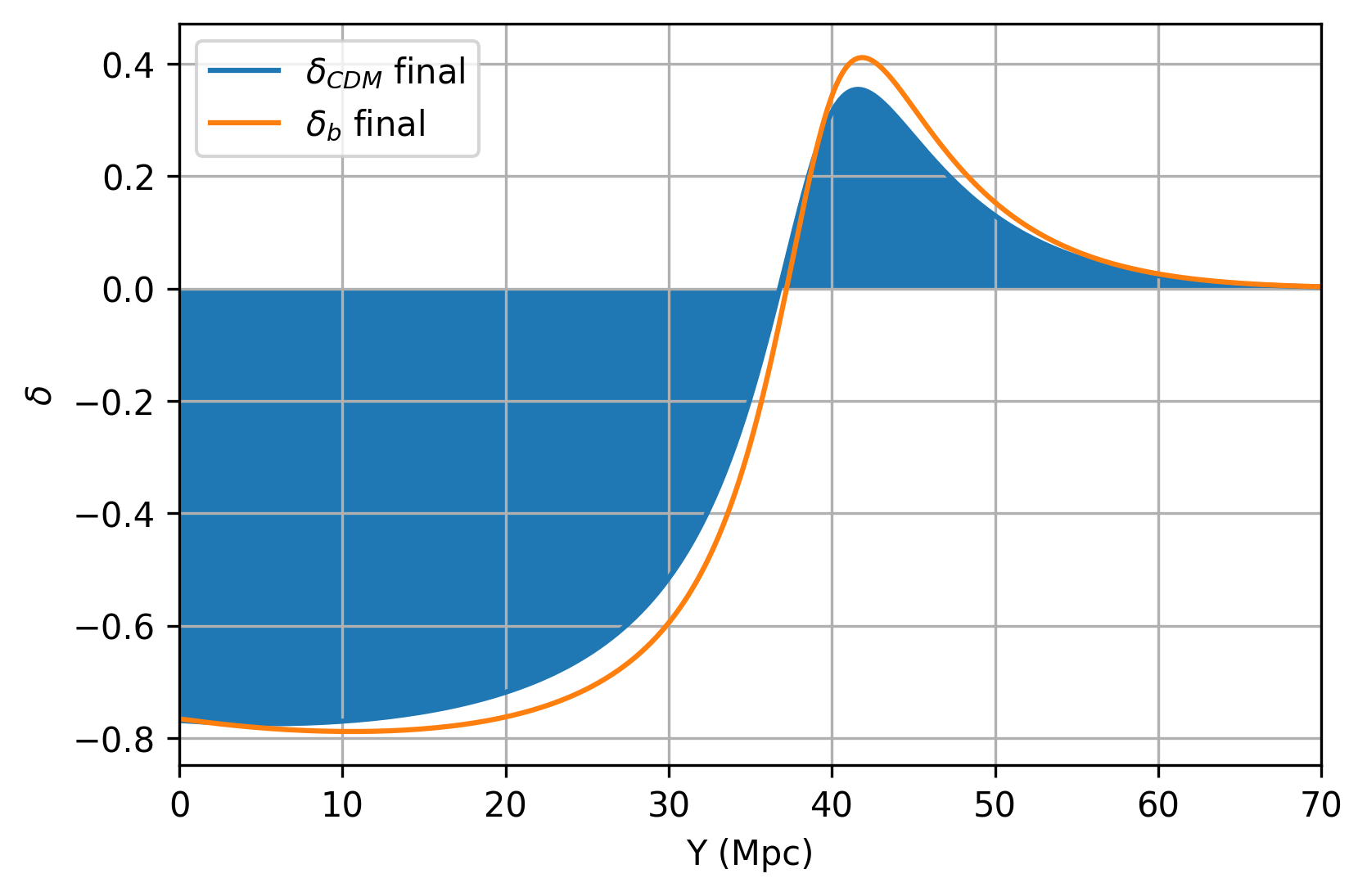}
  \label{Fig:rhoprom}
\end{subfigure}%
\begin{subfigure}{.5\textwidth}
  \centering
  \includegraphics[width=1\linewidth]{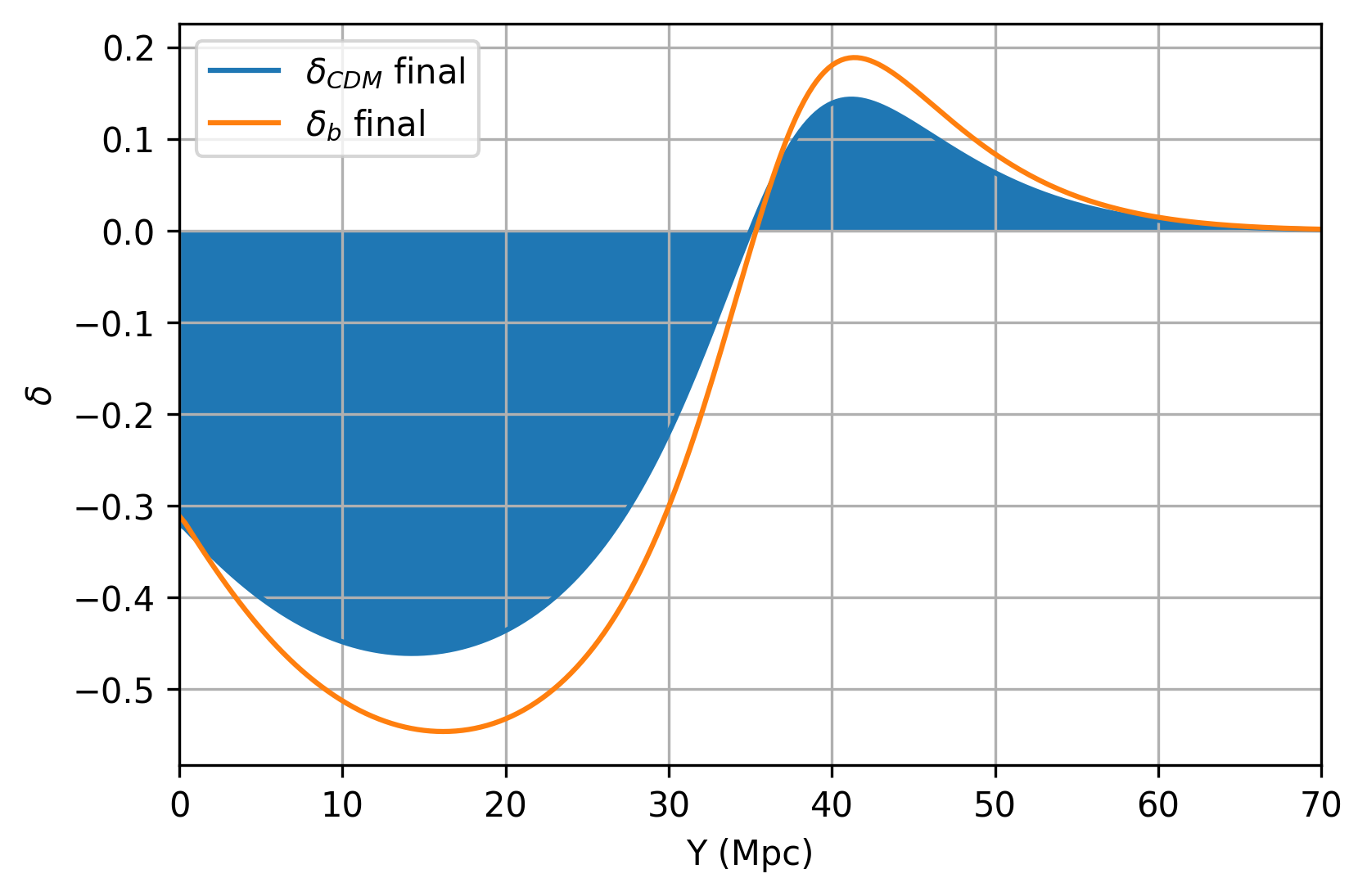}
  \label{Fig:rhopromkc}
\end{subfigure}
\caption{As seen in both graphs, we take the region of negative $\delta$ values and average it over the volume as defined by the metric \eqref{eq:metricainhom} and consider it as the void region while we take the average for the positive values representing the over-dense spherical ridge and consider it as the over-density region. Both of these regions are represented as the filled in regions in both graphs.}
\label{Fig:rhopromT}
\end{figure}

We divide the profiles in two averaged structures  as shown in Fig. \ref{Fig:rhopromT}, representing a void region and the over-dense spherical shell.
From the initial conditions described at the beginning of this section, the resulting growth function $f$ after the numerical evolution for the $k_c=0.05$ case is shown in Fig. \ref{Fig:fOm} within a range of $z \in [0,20]$ for multiple velocities. On the other hand,  the case for $k_c=0.009$ is shown in Fig. \ref{Fig:fOm} for the same range of $z$. They are shown compared to the analytic function obtained from linear perturbation theory. 
%%%%% MOVIDO
 As we can see in Fig. \ref{Fig:cdmbevperf}, the density contrast of both components behave in a similar way  . We can clearly see how the void expands as time passes and even achieves non-linear amplitudes for the density contrast ($\delta_{\rm CDM} \sim -0.8$ at $z=0.1$). Additionally, we observe the formation of a spherical over-dense region surrounding the void region (as represented by the positive values of the density contrast) which also achieves non-linear values ($\delta \sim 0.3$ for the CDM and $\delta \sim 0.4$ for the baryons at $z=0.1$). This is a generic result of the evolution of cosmic voids \cite{1984PThPh..71..938S, 1985ApJS...58....1B, Sheth:2003py}. 
%%%%%%%
We observe in all cases that the behaviour further deviates from the analytic function the higher the maximum velocity is. 
A thing to note is that for the case of no relative velocity between the matter components the CDM and baryonic density contrast and, as a result, their growth function behave exactly in the same manner. As the velocity increases, the discrepancy between the components' behaviour increases. For the total matter density contrast, the effect of the relative velocity is less pronounced as the separate components one. As one can see in both figures, for both curvature cases the numerical growth function gets closer to the analytic one for lower $z$ values for the case of the over-dense ridge excepting the total growth function where the lower $z$ values are where the function differs mostly from the analytic function. For the void region we see the opposite effect, where the lower ranges of redshift are the ones where the numerical and analytic growth functions differ the most with the cases of higher relative velocity exceeding a difference of $40\%$.
For the case of the lower curvature, the deviation from the analytic function is still present but to a lesser extent than for the case of the higher curvature.

\section{Discussion}
\label{Sec:discussion}

In the previous sections we have described the 1+3 splitting of space-time and the Einstein equations. We described how this splitting plus an inhomogeneous model helps  us describe a two-component cosmological system that takes into account a relative velocity between components and uses fully non-linear equations. The effect of a relative velocity between two matter components on a spherical void has been previously studied in \cite{DelgadoGaspar:2018uur} showing the effects this velocity has on the evolution of the void. Similar to that previous work, we developed a numerical system and extended the work to analyze the growth factor $f$. We can see in figures \ref{Fig:fOm} and \ref{Fig:fOmkc} that even the inhomogeneus LTB case with zero relative velocity has a clear distinction from the perturbative case and the relative velocity between baryons and CDM further intensifies this difference. 

%\begin{figure}
%\begin{subfigure}{.5\textwidth}
 % \centering
 % \includegraphics[width=1\linewidth]{curvalta/ferrOD.png}
  %\label{Fig:ferrOD}
%\end{subfigure}%
%\begin{subfigure}{.5\textwidth}
 % \centering
  %\includegraphics[width=1\linewidth]{curvalta/ferrV.png}
  %\label{Fig:ferrV}
%\end{subfigure}
%\begin{subfigure}{.5\textwidth}
 % \centering
  %\includegraphics[width=1\linewidth]{curvalta/ferrbOD.png}
  %\label{Fig:ferrbOD}
%\end{subfigure}%
%\begin{subfigure}{.5\textwidth}
 % \centering
  %\includegraphics[width=1\linewidth]{curvalta/ferrbV.png}
  %\label{Fig:ferrbV}
%\end{subfigure}
%\begin{subfigure}{.5\textwidth}
 % \centering
  %\includegraphics[width=1\linewidth]{curvalta/ferrTOD.png}
 % \label{Fig:ferrTOD}
%\end{subfigure}%
%\begin{subfigure}{.5\textwidth}
 % \centering
  %\includegraphics[width=1\linewidth]{curvalta/ferrTOD.png}
  %\label{Fig:ferrTV}
%\end{subfigure}
%\caption{The figures are arranged in the exact same way as the two previous ones. These graphs display the percentage difference that every graph in Fig. \ref{Fig:fOm} for a range of $z \in [0,2]$ with the added expected ranges to be measured by the DESI colaboration. Note that most of the curves from our numerical results lie outside the DESI ranges.}
%\label{Fig:ferr}
%\end{figure}
%
\begin{figure}[ht]
\begin{subfigure}{.5\textwidth}
  \centering
  \includegraphics[width=1\linewidth]{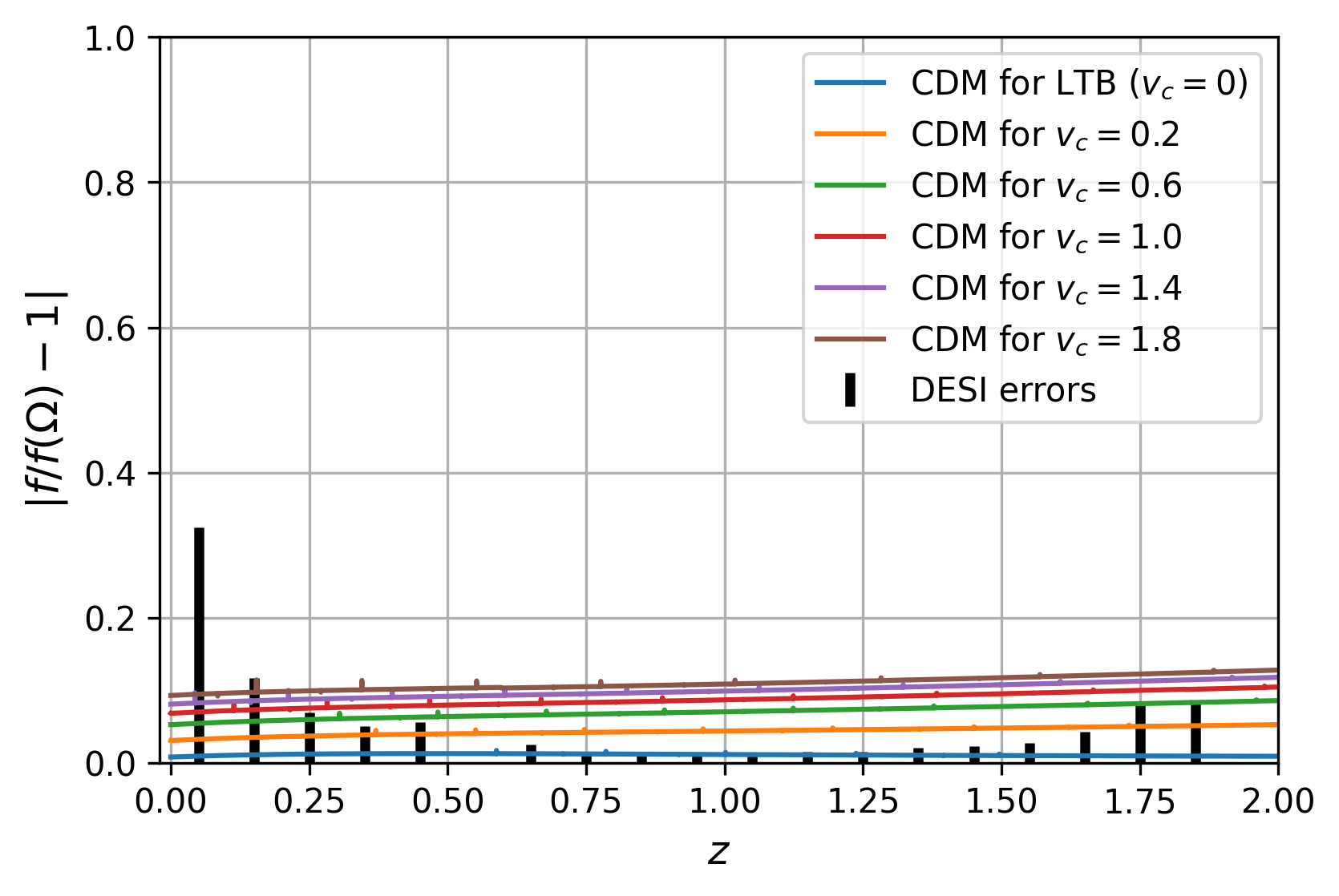}
  \label{Fig:ferrkcOD}
\end{subfigure}%
\begin{subfigure}{.5\textwidth}
  \centering
  \includegraphics[width=1\linewidth]{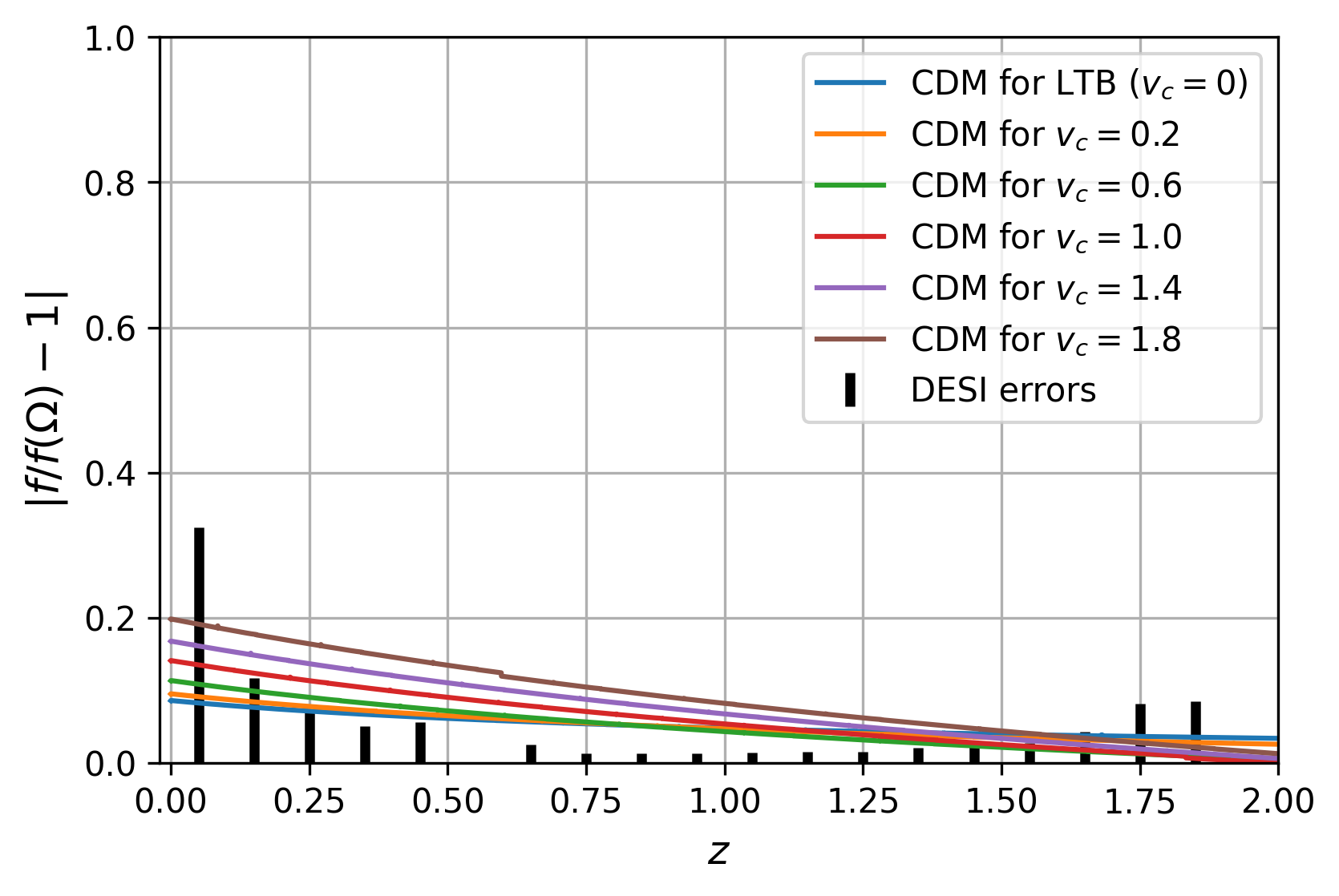}
  \label{Fig:ferrkcV}
\end{subfigure}
\begin{subfigure}{.5\textwidth}
  \centering
  \includegraphics[width=1\linewidth]{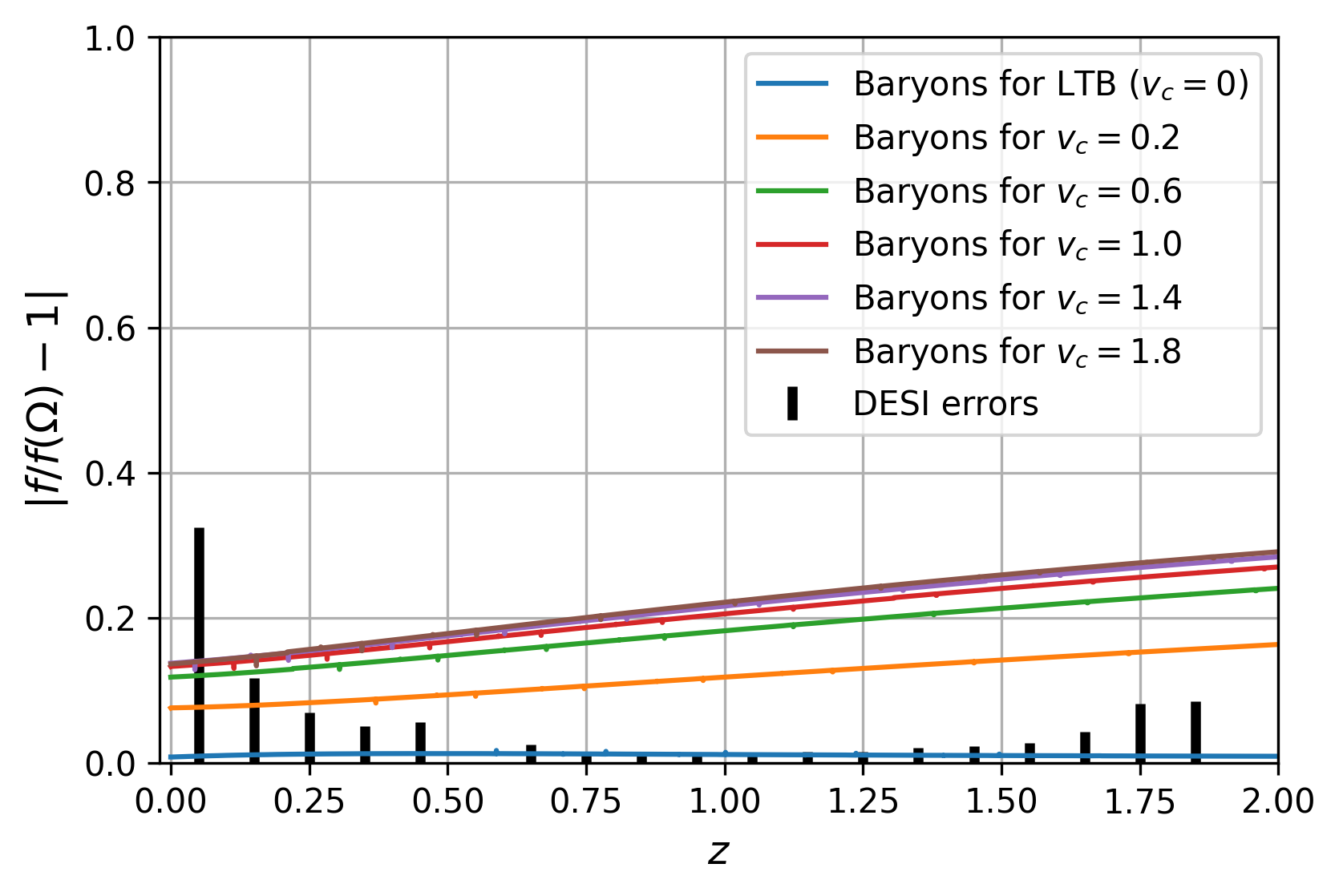}
  \label{Fig:ferrbkcOD}
\end{subfigure}%
\begin{subfigure}{.5\textwidth}
  \centering
  \includegraphics[width=1\linewidth]{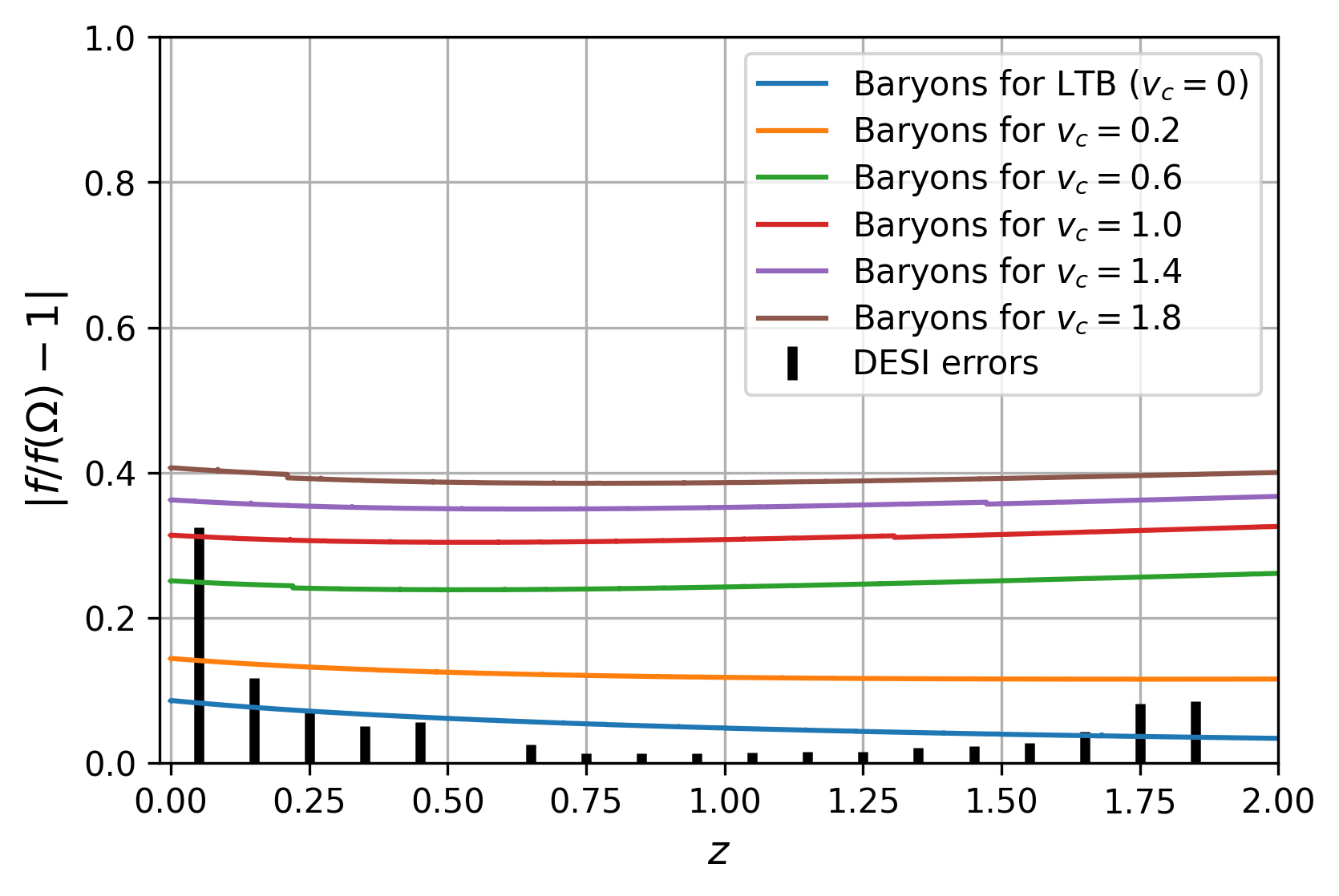}
  \label{Fig:ferrbkcV}
\end{subfigure}
\begin{subfigure}{.5\textwidth}
  \centering
  \includegraphics[width=1\linewidth]{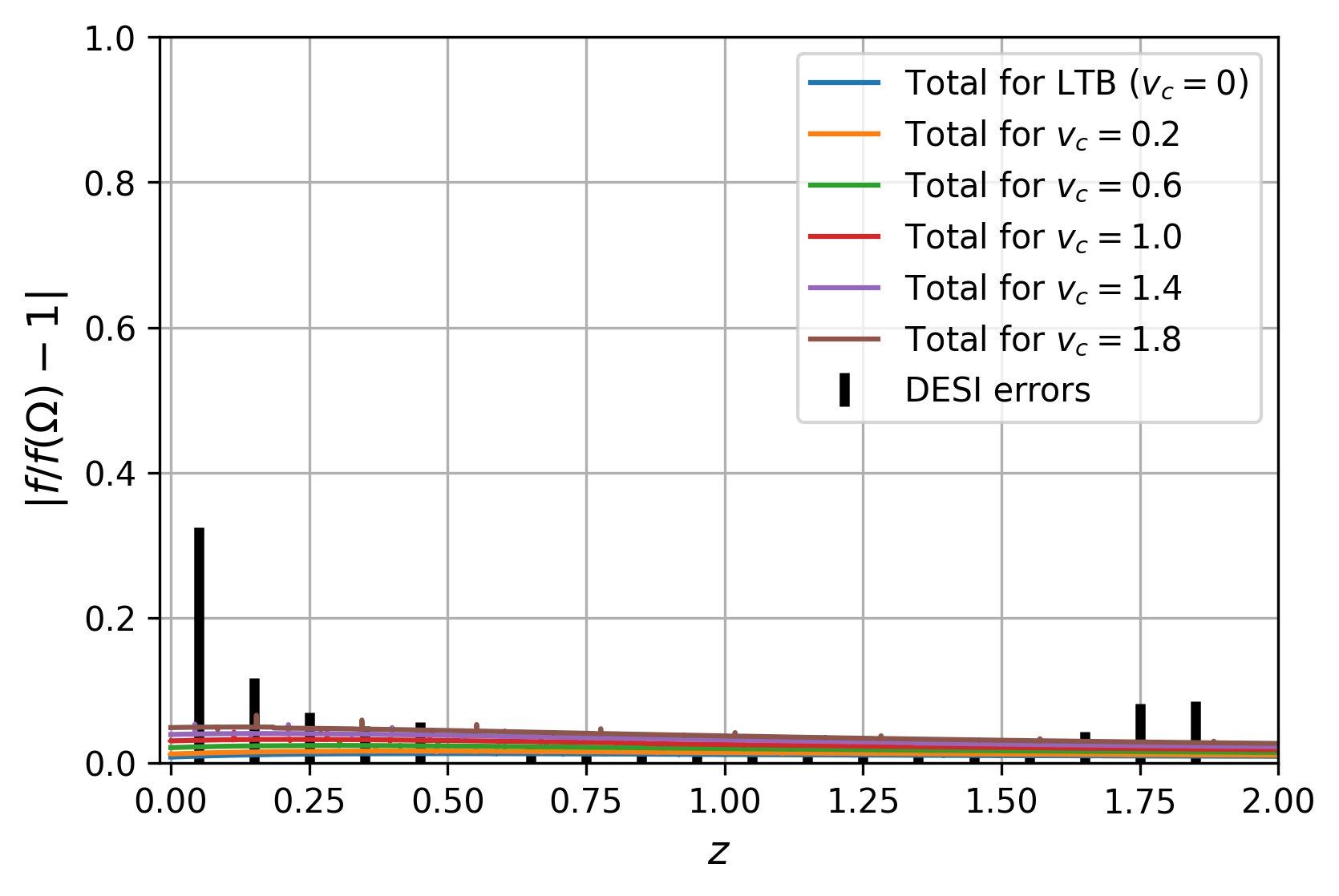}
  \label{Fig:ferrTkcOD}
\end{subfigure}%
\begin{subfigure}{.5\textwidth}
  \centering
  \includegraphics[width=1\linewidth]{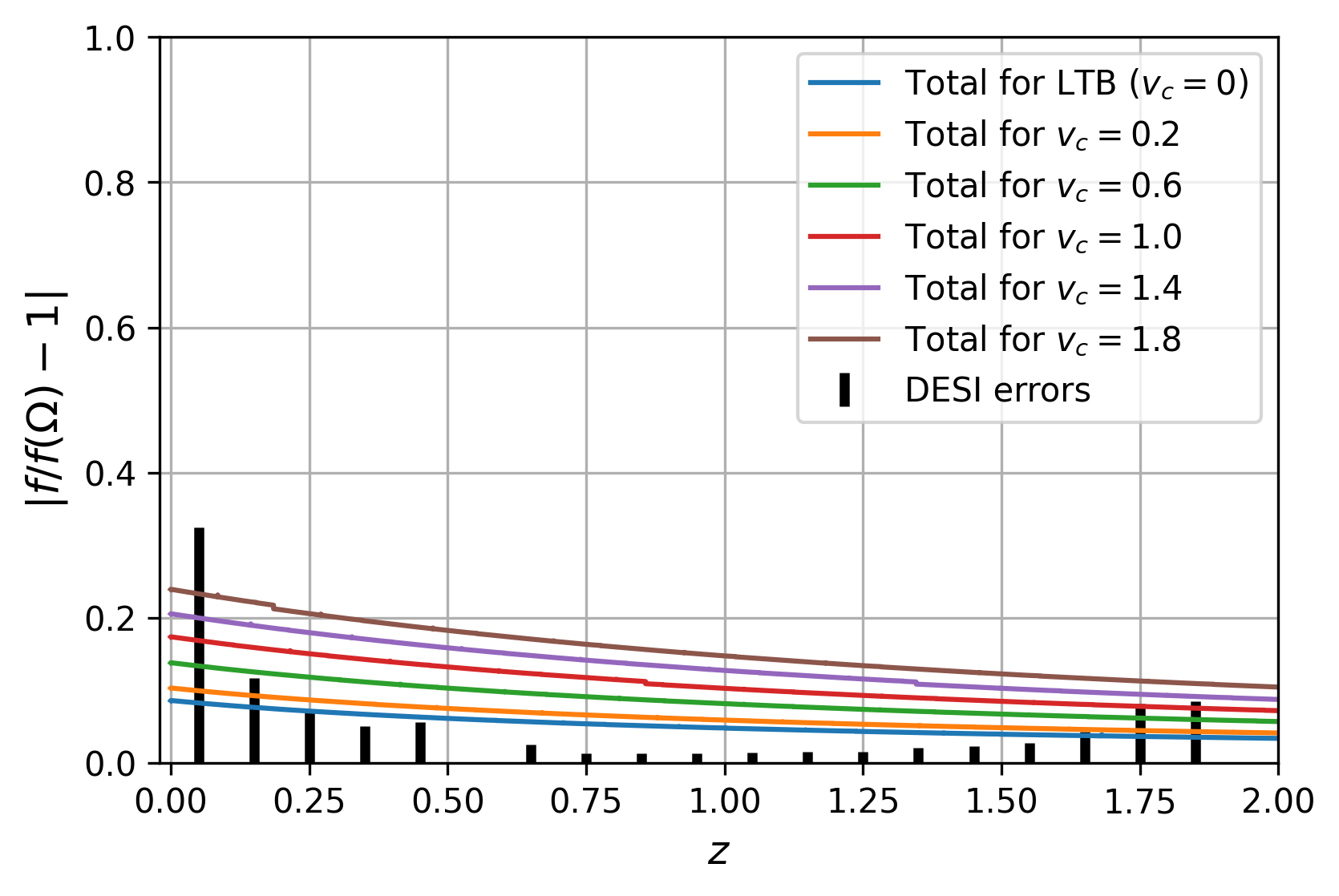}
  \label{Fig:ferrTkcV}
\end{subfigure}
\caption{The figures are arranged in the exact same way as the previous one. These graphs display the percentile difference that every graph in Fig. \ref{Fig:fOmkc} possesses for a range of $z \in [0,2]$ with the added expected ranges to be measured by the DESI colaboration.}
\label{Fig:ferrkc}
\end{figure}
\FloatBarrier
For the following discussion we focus on the low curvature, with $k_c = -0.009$, case. In figure \ref{Fig:ferrkc} we plotted the percentile difference between the analytic growth function of \eqref{eq:fomegahochgam} and our numerical growth functions for the different velocities and the lower curvature compared to the expected margin of meassurement from the DESI collaboration \cite{DESI:2016fyo} plotted in a redshift range of $z \in [0,2]$ for consistency with DESI's report. 

We divert our attention to the over-dense ridge first: as we can see from the left hand-side plots of the figure for both the baryons and CDM  separately the numerical plots lie mostly outside the ranges given by DESI. The one exception is the LTB plot corresponding to no relative velocity between components. All lines corresponding to all the relative velocity amplitudes tested lie within the margin of error of the closest redshift meassurement error available for both fluid components. The effect of having a higher value for $|f/f(\Omega)-1|$ increases with an increase in the relative velocity distribution amplitude; this same effect of increase is seen more pronounced for the case of the baryonic component. Finally when the total density contrast is considered the effects of relative velocity in the growth function are much less pronounced.

By considering now the void region (right hand-side panels) we see a similar effect but more pronounced. Even for the case of CDM and no relative velocity the curve lies outside most of the values of DESI expected errors; while for the baryonic case the two curves corresponding to the higher velocity lie completely outside of all DESI expected errors. One difference between the behaviour of the void and over-dense regions is that the behaviour of the total density contrast for the void lies in between the CDM and baryonic components instead of being the most close to the DESI expected values. 

With all this in mind we conclude that such effects produced by relative velocity between field sources and, to a lesser extent, inhomogeneities is meassurable in future observational collaborations.

There are claims that the non-linearity of LTB models with large density gradients in the supercluster scale is merely an effect of using a comoving gauge that disappears when passing to the conformal Newtonian gauge and considering the values of realistic Newtonian peculiar velocities defined as $v_{\tiny{pec}}=\dot{R}-\bar{H}R$, with $R$ the areal radius of the LTB model (see \cite{Paranjape:2008ai, VanAcoleyen:2008cy}). According to these authors, this supports the claim that the Universe is quasi-Newtonian at these scales \cite{Green:2013yua, Green:2014aga}. These claims are sustained on a gauge transformation from the comoving to the conformal Newtonian gauge applied to the linear approximation of the LTB metric in comoving and non-comoving frames and neglecting higher order terms on $v_{\tiny{pec}} \ll 1$.  However, the definition of these authors of peculiar velocities is completely artificial and ad hoc, while we have defined peculiar velocities properly in terms of the energy flux of an energy-momentum tensor in the context of two dust sources in comoving and non-comoving frames, even if we have also assumed that $v_{\tiny{pec}} \ll 1$. In fact, the reasoning of \cite{Paranjape:2008ai, VanAcoleyen:2008cy} might apply only to the idealized unrealistic case of a single dust source and an ad hoc artificial definition of peculiar velocities. As a contrast, we have considered in this paper a more realistic situation of two sources and a correct relativistic definition of peculiar velocities. Our results clearly shows that the dynamics of the non-comoving dust are relativistic and non-linear,  distinct from the dynamics of dust in a conformal Newtonian gauge of linear perturbations, even if we also assumed peculiar velocities to be Newtonian and an almost LTB metric (\eqref{eq:metricainhom} with $N=1$).  This issue is outside the scope of the present paper, so we will discussed in a separate article.

\appendix

\section{On the definition of the  growth function}
\label{sec:AppA}

 %\edii{
 
In the standard approach, the growth function of cosmic structures is defined in the linear regime, where (considering only the growing mode $D_+$) the density contrast exhibits separate time and spatial dependencies 
\begin{equation}
\delta^L(\mathbf{r},t)=D_+(t)\Delta(\mathbf{r}) \ .
\end{equation}
Then, the growth function is defined as 
\begin{equation}\label{LinGrowthFunc}
    f=\frac{\dot{D}_{+}}{\bar{\bar{H}}D_{+}} = \frac{\mathrm{d} \log D_{+} }{\mathrm{d} \log a}.
\end{equation}
Here $\bar{{H}}$  and $a$ are the Hubble expansion and scale factor of the FLRW background model. From this definition, we can appreciate that under the assumption of linear perturbations, primordial overdensities grow at the same rate everywhere, regardless of the initial matter distribution or curvature perturbation.
This description differs from that of inhomogeneous cosmology, where structures backreact on the gravitational field, and such a background model emerges from averaging the solution on sufficiently large scales.
 
To extend this framework to more general scenarios, we propose the following  definition of the growth function:
\begin{equation}\label{GrowthFuncInh}
    f=\frac{1}{H_{\cD_H}} 
    \frac{ \dotaverage{\delta}}{\average{\delta}}
     =\frac{\mathrm{d} \log \average{\delta} }{\mathrm{d} \log a_{\cD_H}} \ .
\end{equation} 
In the equation above the spatial average is defined as \cite{AverProp-I}\footnote{Despite the fact that we are considering a multi-fluid approach, our observers move with the dark matter and have no acceleration ($N=1$). This entitles us to use the standard average formalism for dust cosmologies. See Ref. \cite{AverProp-II,AverProp-III} for the average properties of general fluids in arbitrary foliations.} 
\begin{equation}
\average{\delta}=\frac{1}{V_\cD} \int_\cD \delta J d^3\mathbf{r} \quad \hbox{with} \quad V_\cD= \int_\cD J d^3\mathbf{r} \ ,
\end{equation}
and the volume element of the spatial hypersurfaces is given by $dV=J d^3\mathbf{r}$. 
The average extends over a compact domain $\cD$ containing the entire structure.  
$a_{\cD_H}$ and  $H_{\cD_H}$ represent the effective scale factor  and averaged expansion rate on the 'scale of homogeneity' (large enough to define the cosmological background)
\begin{equation}\label{EffScalFactor}
a_{\cD_H} (t) = 
\left(\frac{V_{\cD_H}(t)}{V_{\cD_H}(t_\ini)}\right)^{1/3} \ , 
\quad 
H_{\cD_H}=\frac{1}{3}\frac{\dot{V}_{\cD_H}}{V_{\cD_H}}=\frac{\dot{a}_{\cD_H}}{a_{\cD_H}} \ .
\end{equation}

The physical motivation behind \eqref{GrowthFuncInh} becomes more transparent if we consider the following elements. (i) this definition reduces to \eqref{LinGrowthFunc} under the assumptions of linear perturbations ($\average{\delta^{L}} =D_+$ and $a_{\cD_H}\rightarrow a$, the Friedmann scale factor). (ii) The average extends all over the entire structure, resembling the growth of a uniform spherically symmetric overdensity detached from the expansion, as in the so-called ``spherical collapse model'' (for example, see Section 8.2 in \cite{weinberg2008cosmology}). (iii) As thus defined,  $f$ considers the variations in the domain ${\cD}$. Note that a similar definition $ f=\frac{1}{H_{\cD_H}}  \average{\dot{\delta}}/\average{\delta}$ satisfies $(i)$ and $(ii)$ but fails to include the effects of changes in ${\cD}(t)$.

\section{Removing $\sigma_8$ from the DESI report values}
\label{sec:AppB}

In figure \ref{Fig:ferrkc} we present plots for the quantity $|f/f(\Omega)-1|$ and place the DESI values for the expected errors in order to present a comparison. However the DESI collaboration presents results  for the product $f{\sigma_8}$, so we must remove the $\sigma_8$ from the DESI values for a proper comparison with our results. To that end, we start with equation (53) from \cite{Bruni:2013qta}
\begin{equation}
    \frac{D_+}{D_{+*}}=\left( \frac{5}{2f(\Omega_m)+3\Omega_m} \right)\frac{\mathscr{H}}{\mathscr{H}_*}
\end{equation}
Where $\Omega_m=\Omega_{b}+\Omega_{\tiny{CDM}}$. We define $\sigma_8(z)$ and $\Delta \sigma_8(z)$ as
\begin{equation}
    \sigma_8(z) = 0.8111  \frac{D_+}{D_{+*}} \qquad \text{and} \qquad \Delta \sigma_8(z)=0.0060 \frac{D_+}{D_{+*}}.
\end{equation}
Using the values given in tables 2.3 and 2.5 from \cite{DESI:2016fyo} for the errors for $\Delta f\sigma_8(z)$ obtaining for each $z$ given (in the notation of that same reference)
\begin{equation}
    \Delta f \sigma_8(z)=\frac{1}{100}\frac{\sigma_{f_{\sigma_{0.1}}}}{f_{\sigma_{0.1}}}f(\Omega_m(z))\sigma_8(z).
\end{equation}
Finally, we obtain the desired value for $\Delta f(z)$ that we present in figure \ref{Fig:ferrkc}
\begin{equation}
    \Delta f(z)=\frac{\Delta f_{\sigma_8}(z) \sigma_8(z)-f(z)\sigma_8(z)\Delta \sigma_8(z)}{\sigma_8^2(z)}.
\end{equation}

\acknowledgments

JCH acknowledges support from PAPIIT-UNAM project IG102123 ``Laboratorio de Modelos y Datos (LAMOD) para proyectos de Investigación Científica: Censos Astrofísicos" and from CONACyT
Network Project No.~304001 ``Estudio
de campos escalares con aplicaciones en cosmolog\'ia y
astrof\'isica'', and through grant CB-2016-282569. IDG acknowledges the support of the National Science Centre (NCN, Poland) under the Sonata-15 research grant UMO-2019/35/D/ST9/00342. FAP acknowledges financial support from the grants program for postgraduate students of CONACYT as well as MNiSW grant DIR/WK/2018/12.


% Please avoid comments such as "For a review'', "For some examples",
% "and references therein" or move them in the text. In general,
% please leave only references in the bibliography and move all
% accessory text in footnotes.

% Also, please have only one work for each \bibitem.

\end{document}